\documentclass[uplatex,a4paper,prb,cp, preprint, onecolumnt]{revtex4-1}
\usepackage[pdftex]{graphicx}
\usepackage{bm}
\usepackage{amsmath,amssymb}
\usepackage{braket}
\usepackage{ulem}
\usepackage[version=3]{mhchem}
\renewcommand{\Im}{\mathrm{Im}}
\renewcommand{\Re}{\mathrm{Re}}
\renewcommand{\i}{\mathrm{i}}
\newcommand{\Tr}{\mathrm{Tr}}
\newcommand{\unit}[1]{\hat{\boldsymbol{#1}}}

\newcommand{\tro}{{\cal T}}

\newcommand{\pauli}[1]{\hat{\sigma}_{#1}}
\newcommand{\up}{\uparrow}
\newcommand{\dn}{\downarrow}
\newcommand{\thetak}{\theta_{\boldsymbol{k}}}
\newcommand{\phik}{\phi_{\boldsymbol{k}}}
\newcommand{\thetam}{\theta_{\boldsymbol{m}}}
\newcommand{\phim}{\phi_{\boldsymbol{m}}}

\newcommand{\kF}[1]{k_{\mathrm{F} #1}}

\newcommand{\Hs}{\hat{H}_\mathrm{s}}

\newcommand{\Himp}{\hat{H}_\mathrm{imp}}
\newcommand{\Hd}{\hat{H}_{\mathrm{d}}}
\newcommand{\Hhyb}{\hat{H}_\mathrm{hyb}}

\newcommand{\Uss}{\hat{U}_\mathrm{ss}}

\newcommand{\Vsd}{\hat{V}^\mathrm{sd}}
\newcommand{\Vds}{\hat{V}^\mathrm{ds}}
\newcommand{\Vcf}{\hat{V}_\mathrm{cf}}

\newcommand{\Selfenergy}{\Sigma_{\boldsymbol{k}}}

\newcommand{\Gk}{\hat{G}_{\boldsymbol{k}}}
\newcommand{\gk}{\hat{g}_{\boldsymbol{k}}}

\newcommand{\selfenergyss}{\eta_{\mathrm{ss}}}
\newcommand{\Dels}{\Delta_\mathrm{s}}
\newcommand{\Deld}{\Delta_\mathrm{d}}
\newcommand{\Ed}{E_\mathrm{imp}}

\newcommand{\Delc}{\Delta_\mathrm{C}}
\newcommand{\DelT}{\Delta_\mathrm{T}}

\newcommand{\dd}{d^{\dagger}}
\newcommand{\dwidth}{\eta_{\mathrm{d}}}

\newcommand{\sigmaMSHE}{\sigma^{\mathrm{(odd)}}_\perp}
\newcommand{\sigmaASHE}{\sigma^{\mathrm{(even)}}_\perp}
\newcommand{\sigmaSPHE}{\sigma_\parallel^{\mathrm{(odd)}}}

\newcommand{\nimp}{n_{\mathrm{imp}}}
\newcommand{\Nimp}{N_{\mathrm{imp}}}
\newcommand{\cd}{c^{\dagger}}

\newcommand{\EF}{E_{\mathrm{F}}}

\begin{document}
% REVTEX style
\title{Theoretical Study of Extrinsic Spin-current Generation in Ferromagnets Induced by Anisotropic Spin-flip Scattering}
\author{Yuta Yahagi}
\email{yahagi@solid.apph.tohoku.ac.jp}
\affiliation{Department of Applied Physics, Tohoku University, Sendai, Miyagi, Japan.}
\author{Jakub \v{Z}elezn\'{y}}
\affiliation{Institute of Physics, Academy of Sciences of Czech Republic, Prague, Czech Republic}
\author{Daisuke Miura}
\affiliation{Department of Applied Physics, Tohoku University, Sendai, Miyagi, Japan.}
\author{Akimasa Sakuma}
\affiliation{Department of Applied Physics, Tohoku University, Sendai, Miyagi, Japan.}
\begin{abstract}
The spin Hall effect (SHE) and the magnetic spin Hall effect (MSHE) are responsible for electrical spin current generation, which is a key concept of modern spintronics.
We theoretically investigated the spin conductivity induced by spin-dependent s-d scattering in a ferromagnetic 3d alloy model by employing microscopic transport theory based on the Kubo formula. 
We derived a novel extrinsic mechanism that contributes to both the SHE and MSHE. This mechanism can be understood as the contribution from anisotropic (spatial-dependent) spin-flip scattering due to the combination of the orbital-dependent anisotropic shape of s-d hybridization and spin flipping, with the orbital shift caused by spin-orbit interaction with the d-orbitals. We also show that this mechanism is valid under crystal-field splitting among the d-orbitals in either the cubic or tetragonal symmetry.
\end{abstract}
\keywords{Key words}

% JPSJ style
%\title{Title}
%\author{Yuta Yahagi
%	\thanks{Applied Physics, Tohoku University, Sendai, Miyagi, Japan. yahagi@solid.apph.tohoku.ac.jp}, Daisuke Miura, and Akimasa Sakuma}
%\inst{Department of Applied Physics, Tohoku University, Sendai 980-8579, Japan}
%\date{\today}
%\abst{Body of abstract}
%\kword{Key words}

\maketitle

\section{Introduction}
The spin current is a central concept in recent spintronics, and many novel devices driven by the spin current have been proposed, including spin torque magnetic random access memory\cite{Apalkov2016}. Importantly, the use of the spin current in practical applications requires the spin current to be generated efficiently. One of the most practical methods is the spin-transfer effect in ferromagnetic metal bilayers. However, because the spin transfer effect entails not only spin conduction but also charge conduction in principle, it damages the barrier layer, which is a bottleneck in the development of high-density memory\cite{Bhatti2017, Ikegawa2020}.

The spin Hall effect (SHE) and inverse SHE are among the most promising methods for controlling spin currents in next-generation spintronics devices\cite{Sinova2015}.  
SHE is a phenomenon in which a spin current is generated perpendicular to the applied electric field in a material. 
Because the SHE can inject the spin current without a large current flowing through the barrier layer, it is expected to overcome the weakness of the spin transfer effect.
SHE has been studied mainly in non-magnetic heavy metals with a large spin-orbit interaction (SOI), such as \ce{Pt}. 
Recently, however, the SHE in ferromagnetic metals (FMs) has attracted considerable attention\cite{Haney2010, Miao2013, Zimmermann2014, Tian2016, Davidson2020,Qu2020}. 
Some measurements show that the SHE in FMs can generate sufficient spin current to induce magnetization switching with a spin conversion ratio comparable to that in \ce{Pt}\cite{Taniguchi2015, Qin2017, Das2017, Amin2019, Seki2019a, Ma2020}.
The spin swapping effect, a type of SHE that has a perpendicularly polarized spin current, is also enhanced in FMs\cite{Lifshits2009, Pauyac2018}.
Furthermore, spin-current generation owing to the planar-Hall effect (PHE) (a transverse anisotropic magnetoresistance (AMR) effect), rather than the SHE, has also been shown to occur in ferromagnetic metals\cite{Taniguchi2015, Safranski2019, Safranski2020}. 
This PHE-driven spin current (PHE-SC) can also generate a large spin-orbit torque, as demonstrated in recent measurements\cite{Safranski2019,Safranski2020}.
The magnetic spin Hall effect (MSHE) and inverse MSHE, which are new types of charge-spin conversion phenomena, have been reported for a wide range of magnetic materials and have attracted much attention\cite{Zelezny2017, Kimata2019, Naka2019, Mook2020, Gonzalez-Hernandez2020}. 
The MSHE is defined as the time-reversal odd (T-odd) spin-current generation effect; that is, its sign is inverted under a time-reversal operation, unlike the conventional SHE. 
According to this definition, the PHE-SC can be considered as a variant of the MSHE because the PHE-SC has T-odd symmetry as well.
It should be emphasized that the MSHE differs from the spin-polarized current driven by the anomalous Hall effect (AHE), which must be a T-even spin current because of the restriction of the reciprocity theorem among the anomalous Hall conductivities. 
Although the AHE and the associated spin-current generation are driven by Lorentz forces owing to effective magnetic fields, such as Berry curvature, the MSHE is not.
As discussed later, the redistribution of electrons by an applied electric field is essential for the MSHE, rather the effective magnetic fields.

From the perspective of device applications, these magnetization-induced SHE-like effects can have some advantages compared to SHE in non-magnetic materials.
For example, magnetic materials could reduce the use of precious metals. 
Moreover, the spin current in magnetic materials can be easily controlled by modifying the magnetic structure, whereas the spin current in non-magnetic materials is restricted by either the symmetry of the material or the geometry of the device. 
An additional advantage is that the spin current in magnetic materials is less constrained than that in non-magnetic materials because of the symmetry breaking arising from the magnetic structure. 
Magnetic materials can provide polarized spin current parallel to the flow direction, which is preferable for the spin-orbit torque switching of perpendicular magnetization.
%In high-symmetry non-magnetic materials, the spin-polarization is perpendicular to the spin-current flow direction whereas a parallel direction would actually be preferable for spin-orbit torque.

When an electric field $E_j$ is applied to a material, the response spin current $J_i^\mu$ is described as
\begin{equation}\label{eq:SpinCondTensor}
J_i^\mu=\sigma_{ij}^\mu E_j,
\end{equation}
where $\mu$ and $i$ denote the spin polarization direction and flow direction of the spin current, respectively.
The linear response tensor $\sigma_{ij}^\mu$ is referred to as a spin-conductivity tensor.
We decompose $\sigma_{ij}^\mu$ by the time-reversal symmetry, and define the T-even term as the SHE with $\tro\sigma^{\mathrm{SHE}}=+\sigma^{\mathrm{SHE}}$ and the T-odd term as the MSHE with $\tro \sigma^{\mathrm{MSHE}}=-\sigma^{\mathrm{MSHE}}$ 
where $\tro$ denotes the time-reversal operation.
For systems invariant under time-reversal symmetry, such as nonmagnetic materials, the relationship of $\tro\sigma_{ij}^\mu=+\sigma_{ij}^\mu$ holds and the MSHE is forbidden. Therefore, MSHE can only appear in magnetic materials.

The microscopic mechanisms of SHE in non-magnetic materials have been understood with reference to the theory of the anomalous Hall effect (AHE) in ferromagnetic materials.
The major contributions to the SHE are widely believed to be the intrinsic mechanism owing to the Berry curvature of the band structure\cite{Karplus1954} and the extrinsic mechanism due to impurity scattering under the influence of the SOI, as well as skew scattering\cite{Smit1955} and the side jump\cite{Berger1970}.
Hereinafter, we refer to the SHE arising from the intrinsic, skew scattering, and side-jump mechanisms as the conventional SHE.

In magnetic materials, $\sigma_{ij}^\mu$ cannot be fully explained by extending the AHE theory because of the existence of additional contributions.
Recently, several measurements have shown a qualitative disagreement between the anomalous Hall resistivity and the spin Hall resistivity\cite{Omori2019,Koike2020,Hibino2020}.
These results illustrate that the SHE is not always proportional to the AHE, as is sometimes believed, 
and suggest the breakdown of the two current models or the existence of unidentified mechanisms.

The mechanism of the MSHE is not analogous to that of the AHE.
From the viewpoint of Boltzmann transport theory, the MSHE can be expressed by an asymmetric non-equilibrium distribution of the electron spins at the spin-momentum locked Fermi surface, shifted by an external electric field\cite{Zelezny2017}.
Although previous studies on the MSHE have mainly focused on noncollinear antiferromagnets because this effect was initially predicted\cite{Zelezny2017} and observed\cite{Kimata2019} in \ce{Mn3Sn}, the symmetry analysis shows that the MSHE can appear in a wide range of magnetic materials, including typical ferromagnetic metals\cite{Mook2020, Gonzalez-Hernandez2020}. The MSHE was also first predicted for bcc-\ce{Fe}, assuming a simple estimation under the assumption of spin-independent scattering\cite{Zelezny2017}.
In actual ferromagnetic metal systems, however, spin-dependent scattering can play a dominant role. 
The AMR effect is a typical phenomenon in which the contribution from spin-dependent scattering is more dominant than that from spin-independent scattering.
As mentioned above, it has already been partially discussed in the study of the PHE-SC, a subset of the MSHE, under the assumption of the two-current model.
However, spin currents driven by spin-polarized currents are only the secondary effects of a charge current response; thus, it is necessary to proceed beyond the two-current model to completely understand the spin-current response.

\begin{figure}[h]
\centering
\includegraphics[width=0.7\linewidth]{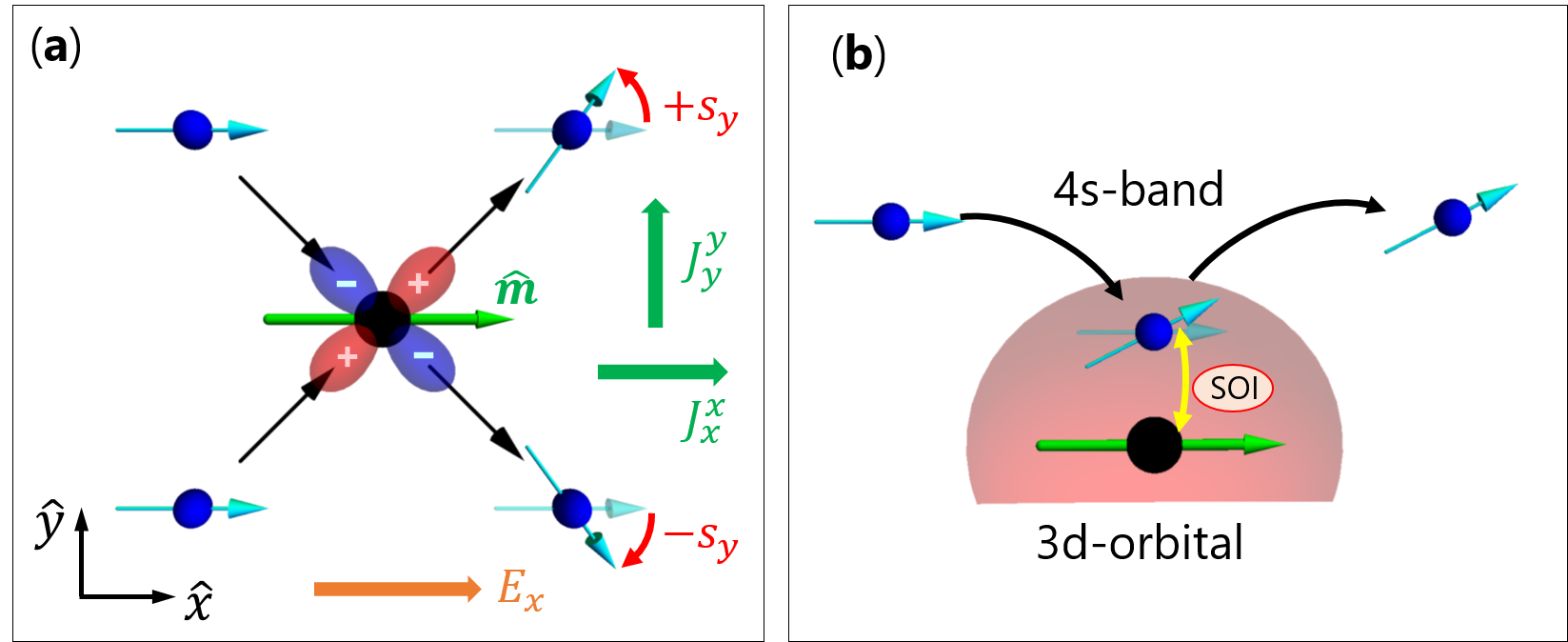}
\caption{(Color online) (a) Real-space view of a spin-current induced by anisotropic spin-flip (ASF) scattering, s-d scattering involving an orbital-selective momentum dependent spin-flip process.
	The blue spheres with arrows represent s-electrons with their spins moving in the direction of the black arrows under the applied electric field $E_x$ and the magnetization $\unit{m}\parallel\unit{x}$.
	The black spheres with red and blue ellipses represent an impurity with a d-orbital. The colors correspond to the sign of the phase.
	For example, an electron moving in the $+y$ direction is more likely to receive a net spin angular momentum $+s_y$ from a momentum-dependent spin-flip, and vice versa for the electrons moving in the $-y$ direction.
	Accordingly, not only is the longitudinal spin current $J_x^x$ associated with the spin-polarized current, but also the transverse spin current $J^y_y$ polarized perpendicular to the magnetization direction.
	(b) Schematic representation of a representative ASF scattering event.
	From a microscopic point of view, a momentum-dependent spin-flip can be realized by a combination of the spin-orbit interaction (SOI) in the d-orbitals and the selection rule for the s-d transition that connects the momentum and d-orbital of the s-electron. 
	This means that the momentum and spin of an electron are coupled via intermediate d-orbitals.}
\label{fig:Schema:ASF}
\end{figure}

The purpose of this work is to systematically analyze the spin currents in ferromagnetic metals with special attention to spin-dependent scattering effects.
We consider a ferromagnetic 3d transition metal dilute alloy model by assuming that the atoms of the minority species are randomly distributed impurities and investigate the impurity s-d scattering in this model. 
The electronic structure is described within the framework of the impurity Anderson model\cite{Anderson1966}, which contains the host-lattice Hamiltonian, impurity Hamiltonian, and their hybridization term.
We identified all the components of the spin current by directly formulating the spin conductivity based on microscopic linear response theory.
Consequently, we found that both the SHE and the MSHE contribute not only from a spin-polarized current but also a distinctive contribution with spin polarization perpendicular to the magnetization direction. 
Such a perpendicularly polarized spin current cannot occur in the two-current model; therefore, this mechanism is separable from the spin-polarized currents driven by either the AHE or PHE.
Our analysis indicated that these contributions arise from anisotropic (spatial-dependent) spin-flip (ASF) scatterings, a type of spin-flip scattering depending on the direction of the momentum of the electron.
Intuitively, the spin current from the ASF scattering can be expressed as spins with different spin angular momenta depending on their direction of motion during the impurity scattering processes, as shown in Fig. \ref{fig:Schema:ASF}.
This scattering process does not occur in an s-s scattering but can occur in the s-d scattering involving an SOI in the d-orbital states and the orbital-selective s-d transition.
The SOI in d-orbitals has two roles: spin-orbit mixing, which causes spin-flipping and orbital splitting as an effective magnetic field.
The orbital-selectivity of the s-d transition re-orientates the spins in a different direction, of which the sign corresponds to the phase of the target d-orbital.
Consequently, the electrons moving in different directions receive different spin angular momenta.
Thus, the ASF scattering induces a momentum imbalance of the non-equilibrium spin distribution, which can drive spin currents polarized both parallel and perpendicular to the magnetization direction.
Hereafter, we refer to the T-even contribution from the ASF as the ASF-SHE and to the T-odd contribution from the ASF as the ASF-MSHE.

The remainder of this paper is structured as follows. In Sect. \ref{sec:model_and_method}, we introduce a one-electron Hamiltonian to describe ferromagnetic dilute alloys in the impurity Anderson model and express the spin conductivity by using the Kubo--Streda formula\cite{Streda1982}. In Sect. \ref{sec:results}, we present the results of perturbation analyses and numerical calculations. In addition, we provide a kinetic picture of ASF-SHE and MSHE.
Finally, we summarize our findings in Sect. \ref{sec:conclusion}.

\section{Model and method}
\label{sec:model_and_method}
\begin{figure}[h]
	\centering
	\includegraphics[width=0.7\linewidth]{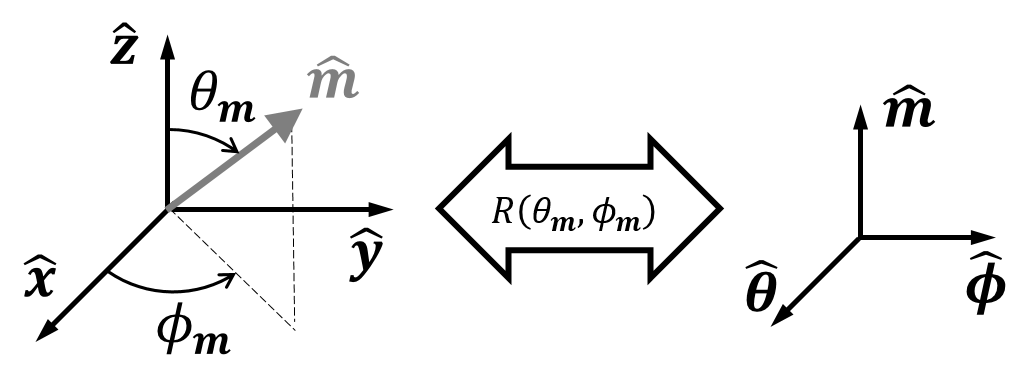}
	\caption{Schematics of the coordination system of the stationary coordinate with the basis $\{\unit{x},\unit{y},\unit{z}\}$ and the rotational coordinate with respect to the magnetization direction $\unit{m}$ with the basis $\{\unit{\theta},\unit{\phi},\unit{m}\}$.}
	\label{fig:Schema:Coordinate2}
\end{figure}

In this study, we used two different coordinate systems, as shown in Fig. \ref{fig:Schema:Coordinate2}. The first is a stationary coordinate system with basis $\{\unit{x},\unit{y},\unit{z}\}$, which we use to represent the spatial coordinates. 
The other is a rotational coordinate system with respect to the magnetization vector with basis $\{\unit{\theta},\unit{\phi},\unit{m}\}$, where we represent the coordinates of the spin space.
In this case, the Pauli matrices are defined as
\begin{equation} \label{eq:Def:PauliMatrix}
\pauli{\theta}=\begin{pmatrix}0&1\\1&0\end{pmatrix},\quad
\pauli{\phi}=\begin{pmatrix}0&-\i\\ \i &0\end{pmatrix},\quad
\pauli{m}=\begin{pmatrix}1&0\\0&-1\end{pmatrix},
\end{equation}
and the spin basis $\{\ket{\up}, \ket{\dn}\}$ is chosen from the eigenfunctions of $\pauli{m}$.
Hereinafter, all the operators represented by the $2\times 2$ matrix in spin space, such as $\hat{A}$, are distinguished by a hat and the $2\times 2$ identity matrix is denoted as $\pauli{0}$.

Focusing on a ferromagnetic 3d transition metal dilute alloy, for simplicity, we consider a downfolded electron band containing 4s- and 3d-bands of the host lattice and localized states of the minority species.
For the host-lattice system, we assume that the 4s-band plays the role of conduction and is described by the electron gas model, whereas the 3d-band plays the role of magnetization and is described by an effective magnetic field under a mean-field approximation.
On the other hand, the localized states of the minority species are regarded as randomly distributed magnetic impurities, and their electronic states are described by localized atomic 3d-orbitals.
In such a situation, the electron Hamiltonian can be described as a multi-orbital impurity Anderson model:
\begin{equation} \label{eq:Def:InitialHamiltonian}
H=\Hs + \Himp +\Hhyb+\Uss,
\end{equation}
where $\Hs$ is the conduction band Hamiltonian of the host-lattice system, $\Himp$ is the atomic 3d-orbital Hamiltonian of the impurity system, and $\Hhyb$ is the hybridization term. $\Uss$ on the right-hand side of Eq. \ref{eq:Def:InitialHamiltonian} represents the s-s scattering through the impurity potential and assumes a spin-independent delta function potential for a sufficiently short distance compared to the mean free path of the conduction electrons.
The conduction band can be described by
\begin{equation}
\label{eq:Def:Hamiltonian_s}
\Hs=\sum_{\boldsymbol{k}} \frac{\hbar^2 k^2}{2m} \cd_{\boldsymbol{k}} c_{\boldsymbol{k}}
-\Dels  \sum_{\boldsymbol{k}} ( \cd_{\boldsymbol{k}}\unit{\sigma} c_{\boldsymbol{k}} ) \cdot \unit{m},
\end{equation}
where $c_{\boldsymbol{k}}=(c_{\boldsymbol{k},\uparrow},c_{\boldsymbol{k},\downarrow})^{\mathrm{T}}$ and $\cd_{\boldsymbol{k}}=(\cd_{\boldsymbol{k},\uparrow},\cd_{\boldsymbol{k},\downarrow})$ denote the creation and annihilation operators of conduction electrons with momentum $\boldsymbol{k}$ and spin $\sigma$. 
$\unit{\sigma}=(\pauli{\theta},\pauli{\phi},\pauli{m})$ is a vector representation of the Pauli matrices in Eq. \eqref{eq:Def:PauliMatrix}.
$\frac{\hbar^2 k^2}{2m}$ represents the kinetic energy of the conduction electron, and $\Dels$ is the strength of the exchange splitting from the ferromagnetic background.
The Hamiltonian of the localized states in a scheme of the Hartree-Fock approximation (HFA) can be described by considering an exchange splitting, an SOI, and a crystal field as follows:
\begin{equation}
\Himp=\sum_{i}^{\Nimp}\sum_{m,m'=-2}^{2}\dd_{i,m'}\left\{
(\Ed\pauli{0}-\Deld\hat{\boldsymbol{\sigma}}\cdot \unit{m})\delta_{m',m}
+\lambda \boldsymbol{l}\cdot\hat{\boldsymbol{\sigma}}
+\Vcf
\right\}d_{i,m}
\end{equation}
\begin{equation}
\begin{split}
\left[\Vcf\right]_{m',m}=&\frac{\Delc}{2}(\delta_{m',\pm 2}\delta_{m,\pm 2}+\delta_{m',\pm 2}\delta_{m,\mp 2}+\delta_{m',\pm 0}\delta_{m,\pm 0})\\
&+\DelT(\delta_{m',\pm 1}\delta_{m,\pm 1}+\delta_{m',\pm 0}\delta_{m,\pm 0}),
\end{split}
\end{equation}
where $d_{i,m}=(d_{i,m,\uparrow},d_{i,m,\downarrow})^{\mathrm{T}}$ and $\dd_{i,m}=(\dd_{i,m,\uparrow},\dd_{i,m,\downarrow})$ denote the creation and annihilation operators of electrons on the 3d-orbital state on the impurity site $i$ with the orbital magnetic quantum number $m$ and spin $\sigma$. $\Ed$ is the energy level center of the impurity state measured from the bottom of the conduction band, and $\Deld$ is the strength of the exchange splitting in the impurity state in the HFA. The electron-electron interaction is treated in the Hartree-Fock approximation and is renormalized into $\Ed$ and $\Deld$.
Note that $\Deld$ includes the influence of the ferromagnetic molecular field created by the 3d-bands of the surrounding host lattice.
$\boldsymbol{l}=(l_{x},l_{y},l_{z})$ denotes each component of the angular momentum operator with $l=2$, and $\lambda$ represents the strength of the SOI.
For the crystal field, we consider the case of cubic or tetragonal crystals and take the crystal axis as $(\unit{a},\unit{b},\unit{c})=(\unit{x},\unit{y},\unit{z})$. $\Delc$ denotes the strength of the cubic field splitting.
Although the strength of the tetragonal-field splitting is not exactly the same for each state, it is represented by a single parameter $\DelT$ for simplicity. 
$\DelT$ denotes the energy difference resulting from tetragonal distortion, that is, the deviation from cubic symmetry.
Because each localized state is independent of the impurity position, $\Himp$ can be divided into the on-site part $\Hd$ and a part representing the impurity configuration, written as
\begin{equation}
\Hd \otimes \left(\sum_{i}^{\Nimp}\ket{i}\bra{i}\right).
\end{equation}

The hybridization term between the conduction band and the localized state is expressed as
\begin{align}
\Hhyb&=\frac{1}{\sqrt{\Omega}}\sum_{i}^{N_\mathrm{imp}} \sum_{\boldsymbol{k},m}\left( \cd_{\boldsymbol{k}} e^{-\i \boldsymbol{k}\cdot \boldsymbol{r}_{i}}\Vsd_{\boldsymbol{k},m} d_{i,m} + H.c. \right),\\
\Vsd_{\boldsymbol{k},m} &= (\Vds_{m,\boldsymbol{k}})^\dagger = 
-\hat{V}(k)Y_{2,m}(\unit{k}),
\end{align}
\begin{equation}
\hat{V}(k)=\begin{pmatrix}V_\up(k)&0\\0&V_\dn(k)\end{pmatrix}
\end{equation}
where $\boldsymbol{r}_{i}$ is the position vector pointing to the impurity center, $\Omega$ is the volume of the system, 
$V_\sigma(k)$ is a radial function with spin $\sigma$, and $Y_{l,m}(\unit{k})$ is a spherical harmonic in the k-space.
Here, we assume that spin mixing via an s-d transition event is negligibly small.

\begin{figure}[h]
	\centering
	\includegraphics[width=0.7\linewidth]{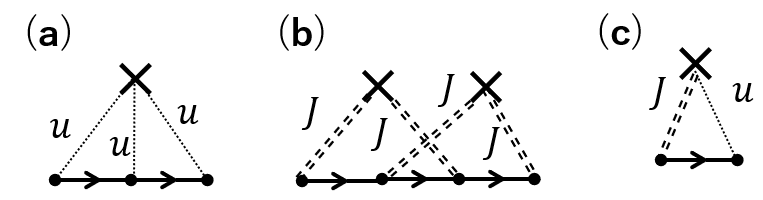}
	\caption{Representative terms of the diagrams excluded from this work. (a) Higher-order Born terms of s-s scattering. (b) Crossing diagrams corresponding to quantum interference between multiple impurities. (c) Onsite interference between s-s and s-d scattering terms.}
	\label{fig:Diagram:Excluded}
\end{figure}

\begin{figure}[h]
	\centering
	\includegraphics[width=0.5\linewidth]{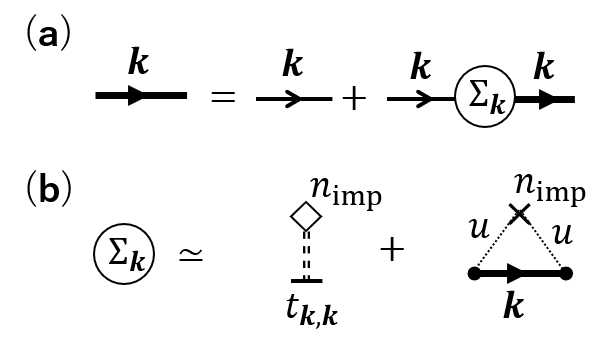}
	\caption{Diagrammatic representations of (a) Dyson's equation for Green's function and (b) the self-energy of the conduction electron. (a) The bold lines represent the clothed Green's function, and the thin line represents the bare one. (b) Self-energy approximated by the configuration-averaged s-d and s-s scattering T-matrix. The double-dashed line represents the s-d scattering, and the dotted line represents the s-s scattering. The point marked by a cross represents a coherent scattering event by a single impurity, whereas the point marked by a diamond represents the scattering by the renormalized effective potential, including all single-site scattering processes.}
	\label{fig:Diagram:GreenFunc}
\end{figure}

To treat the randomness of the impurities, we adopted a scattering T-matrix approach with a method based on Green's function\cite{Altland,mahan2013many}.
The s-s scattering is taken into account within the first Born approximation.
For the s-d scattering, we treat this within the framework of the averaged T-matrix approximation in the dilute limit (DL-ATA), under the assumption that the interference between the impurities is negligible and the impurity concentration is sufficiently dilute.
In addition, we exclude the scattering processes in which the s-s and s-d scattering interfere with each other.
These approximations correspond to the disregard of, for example, the diagrams in Fig. \ref{fig:Diagram:Excluded}. 
In such a case, the Green's function $\Gk$ of the conduction electron and its self-energy $\Selfenergy$ can be expressed as shown in Fig. \ref{fig:Diagram:GreenFunc}
\begin{align}
\Gk^\pm(E)&= \left\{E\pauli{0}-\Hs-\Selfenergy^\pm(E) \right\}^{-1}, \label{eq:Def:GreensFunction}\\ 
\Selfenergy^\pm(E)&=\braket{\hat{T}^\pm_{\boldsymbol{k,k}}}_{\mathrm{conf}}\pm \i \selfenergyss, \label{eq:Def:Selfenergy} 
\end{align}
where $\Gk$ or $\Selfenergy$ with $+(-)$ corresponds to the retarded (advanced) Green's function or self-energy and takes an appropriate sign depending on its own analyticity. 
$\selfenergyss$ is a positive real constant corresponding to the magnitude of s-s scattering and describes the constant spectral broadening of the conduction band.
$\braket{\hat{A}}_{\mathrm{conf}}$ denotes the configurational average of $\hat{A}$ of the impurity; therefore, 
$\braket{\hat{T}_{\boldsymbol{k,k}}}_{\mathrm{conf}}$ represents the configurational average T-matrix of the s-d scattering.
Using $\Nimp \gg 1$, $\braket{\hat{T}_{\boldsymbol{k,k}}}_{\mathrm{conf}}$ is approximately equal to the self-energy from the s-d scattering contribution.

\begin{figure}[h]
	\centering
	\includegraphics[width=0.7\linewidth]{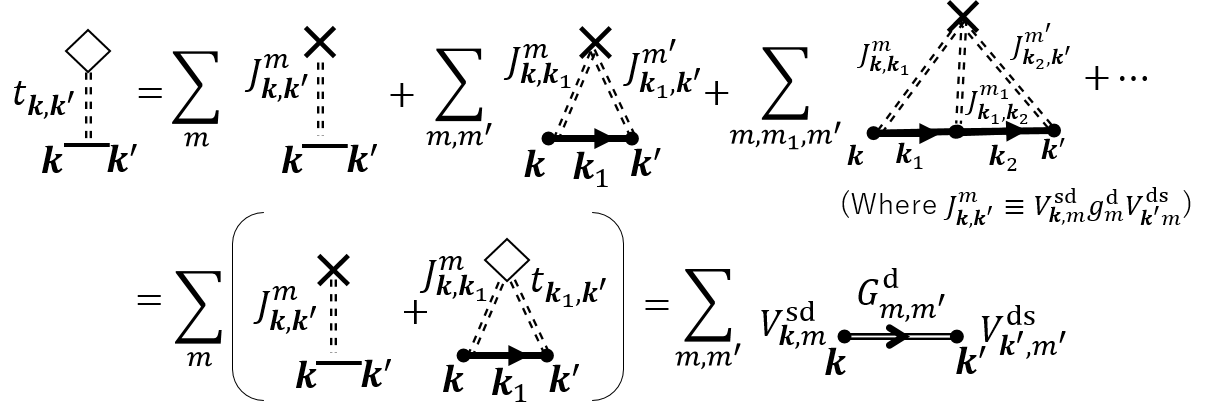}
	\caption{Diagrammatic representation of the single-site s-d scattering T-matrix in the dilute limit averaged T-matrix approximation. 
		Here, $g^{\mathrm{d}}\equiv(E\pauli{0}-\Hd)^{-1}$ represents the bare Green's function of the impurity system, and $G^{\mathrm{d}}\equiv(E\pauli{0}-\Hd-\hat{\varGamma}_{\mathrm{d}})$ represents the clothed function. 
		The suffixes of each factor are omitted for simplicity.
	}
	\label{fig:Diagram:Tmatrix}
\end{figure}

Applying the DL-ATA, we obtain
\begin{equation}\label{eq:Approx:Tmatrix}
\braket{\hat{T}_{\boldsymbol{k,k'}}(E)}_{\mathrm{conf}}\simeq \nimp \hat{t}_{\boldsymbol{k,k'}}(E),
\end{equation}
where $\hat{t}_{\boldsymbol{k,k'}}(E)$ is a T-matrix of single-site scattering, that is, the repeated scattering by the same impurity.
For the impurity Anderson model, $\hat{t}_{\boldsymbol{k,k'}}(E)$ can be derived as 
\begin{align}\label{eq:Approx:SingleSiteTmatrix}
\hat{t}^\pm_{\boldsymbol{k',k}}(E) &=\sum_{m,m'} Y^{\ast}_{2,m'}(\unit{k}')Y_{2,m}(\unit{k})
\hat{V}(k')\left[E\pauli{0}\delta_{m',m}-\Hd-\hat{\varGamma}^\pm_{\mathrm{d}}(E)  \right]^{-1}_{m',m}\hat{V}(k),\\
\because&\left[\hat{\varGamma}^\pm_{\mathrm{d}}(E)\right]_{m',m}=
\sum_{\boldsymbol{k}}\Vds_{m',\boldsymbol{k}}\Gk^\pm(E)\Vsd_{\boldsymbol{k},m}.
\end{align}
The diagrammatic expression of $\hat{t}_{\boldsymbol{k,k'}}(E)$ is shown in Fig. \ref{fig:Diagram:Tmatrix}.
It has almost the same form as the T-matrix of the single impurity Anderson model except that in the case of multiple impurities, it includes a clothed Green's function $\Gk$ instead of a bare one $\gk$.
For simplicity, we approximate $
\hat{\varGamma}^\pm_{\mathrm{d}}(E\pm\i 0) \to \pm\i \dwidth \pauli{0}\delta_{m',m}
$ by introducing a positive real constant, $\dwidth$.
We do not expect this approximation to have a major influence on the results of this work.

Each component of the spin conductivity $\vec{\sigma}_{ij}=(\sigma^m_{ij},\sigma^\theta_{ij},\sigma^\phi_{ij})$ is given by the so-called Kubo-Streda formula\cite{Streda1982}
\begin{align} 
\sigma^\mu_{ij}&=\sigma^{\mu\mathrm{(I)}}_{ij}+\sigma^{\mu\mathrm{(II)}}_{ij}\label{eq:Kubo_Streda0} \\
\sigma^{\mu\mathrm{(I)}}_{ij}&=\frac{\hbar}{4\pi\Omega}\Tr\left\langle J^\mu_i \{\gk^+(\EF) - \gk^-(\EF) \}\hat{J}^0_j\gk^-(\EF)\right. \nonumber \\
&\qquad\qquad\left.-\hat{J}^\mu_i\gk^+(\EF)\hat{J}_j^0\{\gk^+(\EF)-\gk^-(\EF)\}\right\rangle_{\mathrm{conf}} \label{eq:Kubo_Streda1} \\
\sigma^{\mu\mathrm{(II)}}_{ij}&=-\frac{\hbar}{4\pi\Omega}\int_{-\infty}^{\EF}dE
\Tr\left\langle \hat{J}^\mu_i\{\gk^-(E)\}^2\hat{J}_j^0\gk^-(E)
-\hat{J}^\mu_i\gk^-(E)\hat{J}_j^0 \{\gk^-(E)\}^2 \right. \nonumber \\
&\quad\qquad\left.+\hat{J}^\mu_i\gk^+(E)\hat{J}^0_j\{\gk^+(E)\}^2
-\hat{J}^\mu_i\{\gk^+(E)\}^2\hat{J}^0_j\gk^+(E)\right\rangle_{\mathrm{conf}}\label{eq:Kubo_Streda2}.
\end{align}
Here, we assume that the Fermi distribution function of the electron is given by the unit step function, 
which corresponds to the zero-temperature limit.
$\hat{J}^{0}_i$ represents the charge current operator, and 
$\hat{J}^{\mu\neq0}_i\equiv -(\hbar/4e)\{\hat{J}^0_i,\pauli{\mu}\}$ represents an operator of the spin current polarized in the $\mu$-direction.
Here $\{\hat{A},\hat{B}\}$ denotes an anticommutator.
The charge and spin-current operators contain a velocity term derived from the hybridization Hamiltonian, that is, $\partial \Hhyb/ \partial p_i$, which is known to contribute to the side-jump mechanism\cite{Levy1988,Fert2011,Tanaka2009}. 
However, to focus only on the unconventional mechanism, we exclude this term; thus, we set the charge and spin-current operators to
\begin{equation} \label{eq:Def:CurrentOperator}
\hat{J}_i^0=-e\frac{\hbar k_i}{m}\pauli{0},\quad
\hat{J}_i^\mu=\frac{\hbar^2 k_i}{2m}\pauli{\mu}.
\end{equation}

\begin{figure}[h]
	\centering
	\includegraphics[width=0.7\linewidth]{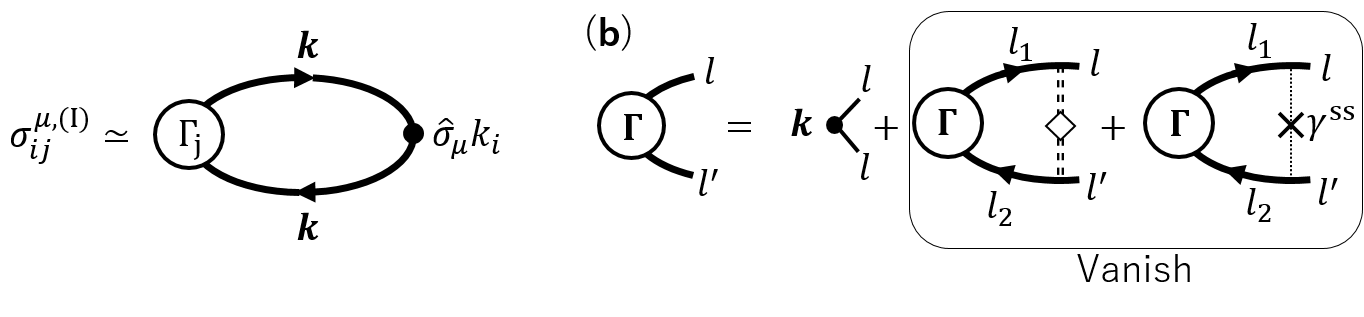}
	\caption{Diagrammatic representation of (a) the current--spin-current correlation function for the Fermi-surface term in the Kubo--Streda formula, and (b) the Bethe-Salpeter equation(BSE) of the current vertex function. The second and third terms on the right-hand side of the BSE vanish owing to the symmetry consideration of the s-d and s-s scattering as $\Gamma_{l,l'}(-\boldsymbol{k})=-\Gamma_{l,l'}(\boldsymbol{k})$, which leads to $\sum_{\boldsymbol{k}}\Gamma_{l,l'}(\boldsymbol{k})f(k)=0$. Therefore, the current vertex function can be replaced by a bare current operator.}
	\label{fig:Diagram:ResponseFunc}
\end{figure}

The current vertex correction for the configuration average is considered only for the terms containing $\braket{\hat{G}^\pm \hat{J}^{0}_i \hat{G}^\mp}_{\mathrm{conf}}$ because these terms are typically much more dominant in metals than in other terms such as $\braket{\hat{G}^\pm \hat{J}^{0}_i \hat{G}^\pm}_{\mathrm{conf}}$.
Therefore, in this work, the vertex correction appears for $\sigma^{\mu\mathrm{(I)}}_{ij}$, and the Bethe-Salpeter equation (BSE) for the current vertex $-(e\hbar/m)\hat{\boldsymbol{\Gamma}}_{\boldsymbol{k,k'}}$ is expressed as
\begin{equation}\label{eq:Def:Bethe-SalpeterEq}
\begin{split}
\left[\hat{\Gamma}_{\boldsymbol{k,k'}}^{+-}\right]_{\alpha,\beta}=&\delta_{\boldsymbol{k',k}}\delta_{\beta,\alpha}\boldsymbol{k},\\ 
&+\nimp^2\sum_{\{\boldsymbol{k}_i\},\{s_i\}}[\hat{\Gamma}_{\boldsymbol{k}_1,\boldsymbol{k}_2}^{+-}]_{s_1,s_2} 
[\hat{G}^+_{\boldsymbol{k}_1}]_{s_1,s_3}[\hat{G}^-_{\boldsymbol{k}_2}]_{s_2,s_4}
[\hat{t}^+_{\boldsymbol{k}_1,\boldsymbol{k}}]_{s_3,\alpha}[\hat{t}^-_{\boldsymbol{k}_2,\boldsymbol{k'}}]_{s_4,\beta},\\
&+\gamma^{\mathrm{ss}}\sum_{\boldsymbol{k},\boldsymbol{k'},s,s'}[\hat{\Gamma}_{\boldsymbol{k},\boldsymbol{k'}}^{+-}]_{s,s'} 
[\hat{G}^+_{\boldsymbol{k}}]_{s,\alpha}[\hat{G}^-_{\boldsymbol{k'}}]_{s',\beta},
\end{split}
\end{equation}
with the abbreviation for $(E)$ in each term. The diagrammatic expressions of $\sigma^{\mu\mathrm{(I)}}_{ij}$ and BSE are 
shown in Fig. \ref{fig:Diagram:ResponseFunc}.
The second and third terms on the right-hand side of Eqs. \eqref{eq:Def:Bethe-SalpeterEq} correspond to the vertex corrections from s-d and s-s scattering, respectively. The s-s scattering vertex function $\gamma^{\mathrm{ss}}$ is a constant.
The symmetry considerations show that these vertex corrections vanish after taking the momentum integration
because both integrands from the s-d and s-s scattering are odd functions with respect to $\boldsymbol{k}$ as $\boldsymbol{k}Y^{\ast}_{l,m}(\unit{k'})Y_{l,m}(\unit{k})$ for $(l=0,2)$.
\footnote{
Only $\hat{t}_{\boldsymbol{k'},\boldsymbol{k}}$, and $\Gk$ depend on the momentum direction $\unit{k}$, and the other terms are isotropic with respect to the momentum space. 
Obviously, $\hat{t}_{\boldsymbol{k'},\boldsymbol{k}} \propto  Y^{\ast}_{2,m'}(\unit{k'})Y_{2,m}(\unit{k})$ is an even function with respect to $\boldsymbol{k}$ and $\boldsymbol{k'}$ as $Y_{lm}(\unit{k})$ becomes an even function when $l$ is an even number.
For $\Gk$, we set the isotropic terms as $\gk'\equiv (\gk^{-1}\pm\i \selfenergyss)^{-1}$, and then expand $\Gk$ with respect to $\hat{t}_{\boldsymbol{k},\boldsymbol{k}}$ as
$$
\Gk=\gk' + \nimp \gk' \hat{t}_{\boldsymbol{k,k}}\gk' + \nimp^2 \gk' \hat{t}_{\boldsymbol{k,k}}\gk' \hat{t}_{\boldsymbol{k,k}}\gk' + \cdots.
$$
The n-th general term is proportional to $\{Y^{\ast}_{2,m'}(\unit{k})Y_{2,m}(\unit{k})\}^n$; therefore, $\Gk$ is also an even function with respect to $\boldsymbol{k}$.
Thus, the general terms of the response function vanish as the integration of an odd function, except for the first term ($\delta_{\boldsymbol{k',k}}\delta_{\beta,\alpha}\boldsymbol{k}$).
}

Eventually, the configurational averaged response function can be obtained by replacing $\gk\to\Gk$ and is derived as
\begin{align}
\sigma_{ij}^{\mu,\mathrm{(I)}}&=A_\mu \left(\frac{\hbar^2}{m}\right)^2\frac{1}{(2\pi)^3}
\int d^3\boldsymbol{k}k_i k_j \Re \Tr_{\sigma} \pauli{\mu}
\left\{\Gk^+(\EF)\Gk^-(\EF) - \Gk^+(\EF)\Gk^+(\EF)\right\},\label{eq:Approx:KuboStreda1}\\
\sigma_{ij}^{\mu,\mathrm{(II)}}&=A_\mu \left(\frac{\hbar^2}{m}\right)^2\frac{1}{(2\pi)^3}
\int d^3\boldsymbol{k}\int_{-\infty}^{\EF}dE  k_i k_j \Re \Tr_{\sigma} \pauli{\mu}
\left[\Gk^+(E),\frac{\partial \Gk^+(E)}{\partial E}\right].\label{eq:Approx:KuboStreda2}
\end{align}
$A_\mu$ is the coefficient of charge conductivity $A_0=e^2/h$ for $\mu=0$, and the coefficient of spin conductivity $A_\mu=-e/4\pi$ otherwise. 
Both $\sigma_{ij}^{\mu,\mathrm{(I)}}$ and $\sigma_{ij}^{\mu,\mathrm{(II)}}$ do not change their sign under the permutation of the direction ($i\leftrightarrow j$); therefore, the resultant spin conductivity tensor is a symmetric tensor.
Note that the skew scattering contribution does not appear in this expression because Eqs. \eqref{eq:Approx:KuboStreda1} and \eqref{eq:Approx:KuboStreda2} consists only of the coherent term owing to the absence of the vertex correction.
Skew scattering as a result of vertex correction is known to appear when considering the scattering between states with different parity, such as d-p scattering\cite{Fert1973,Fert1976,Levy1988,Fert2011,Takata2017}.
The side-jump contribution does not appear because it requires relativistic corrections in the current operators\cite{Crepieux2001}, which is not considered in Eq. \eqref{eq:Def:CurrentOperator}.
In addition, the intrinsic mechanism disappears in Eqs. \eqref{eq:Approx:KuboStreda1} and \eqref{eq:Approx:KuboStreda2} because the conduction band is assumed to consist only of s-electrons, where the SOI does not take any effect unless the d-orbital SOI of the impurity is induced extrinsically.
Thus, we emphasize that all of the conventional mechanisms (intrinsic, skew scattering, and side-jump) are excluded from Eqs. \eqref{eq:Approx:KuboStreda1} and \eqref{eq:Approx:KuboStreda2}.
In addition, the spin swapping effects are also excluded because they have the same origins that are associated with the conventional mechanisms of the SHE\cite{Lifshits2009,Sadjina2012,Pauyac2018}.
Hence all the contributions that we study in this work arise from novel mechanisms.

\section{Calculation Results}
\label{sec:results}
\subsection{Perturbative analysis}
We derive an analytical expression for $\sigma^\mu_{ij}$ by taking a perturbation expansion with respect to the SOI up to the 2nd order.
For simplicity, the crystal fields are disregarded in this section, thus $\Delc=\DelT=0$.
The non-perturbative terms of the d-orbital Green's function are represented as
\begin{equation}
\hat{\Upsilon}^\pm(E)= \begin{pmatrix}
\Upsilon^\pm_\up(E) & 0 \\ 0& \Upsilon^\pm_\dn(E) 
\end{pmatrix}\equiv\begin{pmatrix}
(E-\Ed+\Deld\mp\i \dwidth)^{-1} & 0 \\ 0& (E-\Ed-\Deld\mp\i \dwidth)^{-1}
\end{pmatrix}.
\end{equation}
Hereinafter, their arguments are abbreviated.

We focus on the $\sigma_{yx}^\mu$ component of the spin-conductivity tensor because the tensor is symmetric. Thus, $\sigma_{xy}^\mu=\sigma_{yx}^\mu$, 
Furthermore, other geometries, such as $\sigma_{zy}^\mu$, can be replicated by rotating the coordinate axis.
Here, we also focus on the transverse spin conductivity; the longitudinal conductivity is given in Appendix \ref{sec:Appendix:Longitudinal}.
In Appendix \ref{sec:Appendix:Derivation_Perturbation}, we derive an analytical expression for the spin conductivity.
Consequently, we obtain the transverse spin conductivity $\sigma_{yx}^\mu$ as
\begin{align}
\label{eq:Result:SPHE}
\sigma_{yx}^m=&-\sigmaSPHE \sin^2\thetam \sin 2\phim
,\\
\label{eq:Result:ASHE}
\sigma_{yx}^\theta=&
\frac{1}{2}\sigmaMSHE \sin 2\thetam \sin 2 \phim-\sigmaASHE\sin \thetam \cos 2\phim
,\\
\label{eq:Result:MSHE}
\sigma_{yx}^\phi=&
-\sigmaMSHE\sin \thetam \cos 2\phim -\frac{1}{2}\sigmaASHE\sin 2\thetam \sin 2 \phim,
\end{align}
with the coefficients $\sigmaSPHE,\sigmaMSHE$, and $\sigmaASHE$ including a radial integral in the momentum space.  
The subscripts $\parallel$ and $\perp$ denote the spin polarization direction of the spin current parallel or perpendicular to the magnetization direction, respectively. 
$(\mathrm{even})$ and $(\mathrm{odd})$ denote the even and odd parts with respect to the permutation of the spin basis, which corresponds to the time-reversal symmetry. 
Eqs. \eqref{eq:Result:ASHE} and \eqref{eq:Result:MSHE} indicate that there are both contributions to the SHE and to the MSHE.
Interestingly, all contributions in Eqs. \eqref{eq:Result:SPHE}-\eqref{eq:Result:MSHE} disappear in $\thetam=0$ unlike the contribution from the spin-polarized current driven by the AHE, which is maximized in this angle.

Because we consider dissipative conduction in metallic bands, the terms including $\Gk^+(\EF)\Gk^-(\EF)$ would be expected to dominantly contribute to conduction.
Assuming $\selfenergyss=\frac{\hbar}{2\tau_0} \ll \EF$, we ultimately have
\begin{align}
\sigmaSPHE \simeq&\frac{3\nimp \lambda^2}{2e}\left\{
\sigma_\up \tau_\up V_\up^2(\kF{s}) \Im\left[(\Upsilon_\up^+-\Upsilon_\dn^+)(\Upsilon_\up^+)^2\right] \right. \nonumber \\
&+\left.
\sigma_\dn \tau_\dn V_\dn^2(\kF{s}) \Im\left[(\Upsilon_\up^+-\Upsilon_\dn^+)(\Upsilon_\dn^+)^2\right]
\right\},\label{eq:Approx:SPHE}\\
\sigmaASHE \simeq&-\frac{\hbar}{2e}\frac{3\nimp\lambda^2}{8\pi(E_\up-E_\dn)}
\left\{
\sigma_\up V_\up(\kF{\dn})V_\dn(\kF{\dn}) +\sigma_\dn V_\up(\kF{\up})V_\dn(\kF{\up})
\right\}
\Im\left[(\Upsilon_\up^+-\Upsilon_\dn^+)\Upsilon_\up^+ \Upsilon_\dn^+\right],
\label{eq:Approx:ASHE}\\
\sigmaMSHE\simeq&-\frac{\hbar}{2e}\frac{3\nimp\lambda^2}{8(E_\up-E_\dn)}
\left\{
\sigma_\up V_\up(\kF{\dn})V_\dn(\kF{\dn})- \sigma_\dn V_\up(\kF{\up})V_\dn(\kF{\up})
\right\}
\Re\left[(\Upsilon_\up^+-\Upsilon_\dn^+)\Upsilon_\up^+ \Upsilon_\dn^+\right],
\label{eq:Approx:MSHE}
\end{align}
where $\sigma_\sigma\equiv e^2 n_\sigma \tau_0/m$ represents the nonperturbative conductivity of the spin-$\sigma$ band with the Fermi momentum $\kF{\sigma}$ and the carrier concentration of the Fermi gas $n_\sigma=\frac{\kF{\sigma}^3}{6\pi^2}$.
Because all the expressions have $(\Upsilon_\up^+-\Upsilon_\dn^+)$, these terms never appear for a nonmagnetic impurity.
$\sigmaSPHE$ in Eq. \eqref{eq:Result:SPHE} represents the spin-conduction polarized parallel to the magnetization.
This corresponds to the PHE-SC in the two-current model limit, which treats each spin band independently\cite{Taniguchi2015}.
Note that this picture can be valid only for systems in which spin band mixing is sufficiently small.
In contrast to $\sigmaSPHE$ , $\sigmaASHE$, and $\sigmaMSHE$ in Eqs. \eqref{eq:Approx:ASHE} and \eqref{eq:Approx:MSHE} represent the spin conduction polarized along a direction perpendicular to the magnetization.

\begin{figure}
\centering
\includegraphics[width=0.7\linewidth]{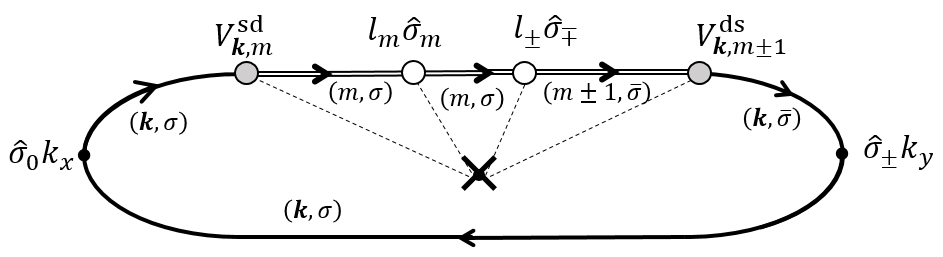}
\caption{Diagrammatic representation of an s-d scattering process contributing both to the magnetic and the spin Hall effect, which corresponds directly to Fig. \ref{fig:Schema:ASF}(b). The solid and double lines represent the electrons of the conduction bands and impurity d-orbital states, respectively. The gray points represent s-d hybridization, and the white point represents the spin-orbit interaction on the d-orbital states. }
\label{fig:Diagram:ElementaryProcess}
\end{figure}

To determine the microscopic mechanism of both $\sigmaASHE$ and $\sigmaMSHE$, we analyzed the elementary processes of scattering, as shown in Fig. \ref{fig:Diagram:ElementaryProcess}.
Here, we use the ladder operators $\pauli{\pm}$ instead of $\pauli{\theta}$ and $\pauli{\phi}$.
This process is referred to as anisotropic spin-flip (ASF) scattering, highlighting its two distinctive features: spin-flipping and anisotropic ($\unit{k}$-dependent) hybridization strength.
In ASF, the electrons are observed as the spin current of $\pauli{\pm}$ after experiencing s-d hybridization and propagation in the d-state with the SOI.
The SOI plays two important roles in ASF as orbital splitting ($l_m\pauli{m}$) and orbital elevation with spin-flipping ($l_\pm\pauli{\mp}$).
The spin-flipping leads to a perpendicular-polarized spin current, which includes both the T-even term and the T-odd term, as shown in Eqs. \eqref{eq:Result:ASHE} and \eqref{eq:Result:MSHE}.
Moreover, because of the orbital elevation, the incoming and outgoing d-orbital states can be different, such as $\Vsd_m \Vds_{m'\neq m}$.
Because $\Vsd\propto Y_{2,m}(\unit{k})$ has an $m$-dependent anisotropic shape, a mixture of  $\Vsd_m$ and $\Vds_{m' \neq m}$ is required for the spin conductivity to have not only the longitudinal component ($\sigma_{ii}^\mu$), but also the transverse component ($\sigma^\mu_{ij}$). 
The orbital splitting $l_m$ is required to avoid the cancelation of $\Vsd_{m} \Vds_{m\pm1}$ and $\Vsd_{-m} \Vds_{-(m\pm1)}$; therefore, the ASF-SHE/MSHE should be obtained from the second- or higher-order perturbation with respect to the SOI.  
For example, when the incoming state $m=0$ and outgoing state $m'=+1$ in the case of $\unit{m}\parallel\unit{x} \quad (\thetam=\pi/2,\phim=0)$,
the finite transverse spin-current contribution $\sigma_{yx}^\pm$ can be obtained as
\footnote{We use the relation of 
	$Y_{2,0}(\hat{k})\propto (3\hat{k}_m^2-1)\propto (3\hat{k}_x^2-1)$ 
	and $Y_{2,\pm1}(\hat{k})\propto \hat{k}_m(\hat{k}_\theta\pm\i\hat{k}_\phi)\propto \mp\hat{k}_x (-\hat{k}_z\pm\i\hat{k}_y)$ when $\hat{m}=\hat{x}$.}
\begin{equation}
\sigma_{yx}^\pm \propto \int d\boldsymbol{k}\hat{k}_x \hat{k}_y (3\hat{k}_x^2-1) \left\{ \hat{k}_x(\hat{k}_z-\i \hat{k}_y) \right\}\neq 0.
\end{equation}

From Eq. \eqref{eq:Result:ASHE} or Eq. \eqref{eq:Approx:MSHE}, we can obtain a scaling law for the spin Hall resistivity $\rho^{\mathrm{SHE(MSHE)}} \equiv  \sigma^{\mathrm{even(odd)}}_\perp/(\sigma_\up+\sigma_\dn)^2$ with respect to the longitudinal charge resistivity $\rho_{xx}$ as
\begin{equation}
\rho^{\mathrm{SHE}}\propto \rho_{xx},\quad \rho^{\mathrm{MSHE}}\propto P(\EF)\rho_{xx},
\end{equation}
where $P(\EF)\equiv(\sigma_\up-\sigma_\dn)/(\sigma_\up+\sigma_\dn)$ is the spin-polarization ratio at the Fermi energy.
Both $\rho^{\mathrm{SHE}}$ and $\rho^{\mathrm{MSHE}}$ scale linearly with respect to $\rho_{xx}$, which is the same as the skew scattering mechanism of the conventional SHE.

Finally, we note that the results of Eqs. \eqref{eq:Result:SPHE}, \eqref{eq:Result:ASHE}, and \eqref{eq:Result:MSHE} are represented in the rotational coordinate system for the spin space. 
Appendix \ref{sec:Appendix:Stationary} provides the spin conductivities transformed into the stationary coordinate system.

\subsection{Kinetic view of the unconventional SHE}
\begin{figure}
	\centering
	\includegraphics[width=\linewidth]{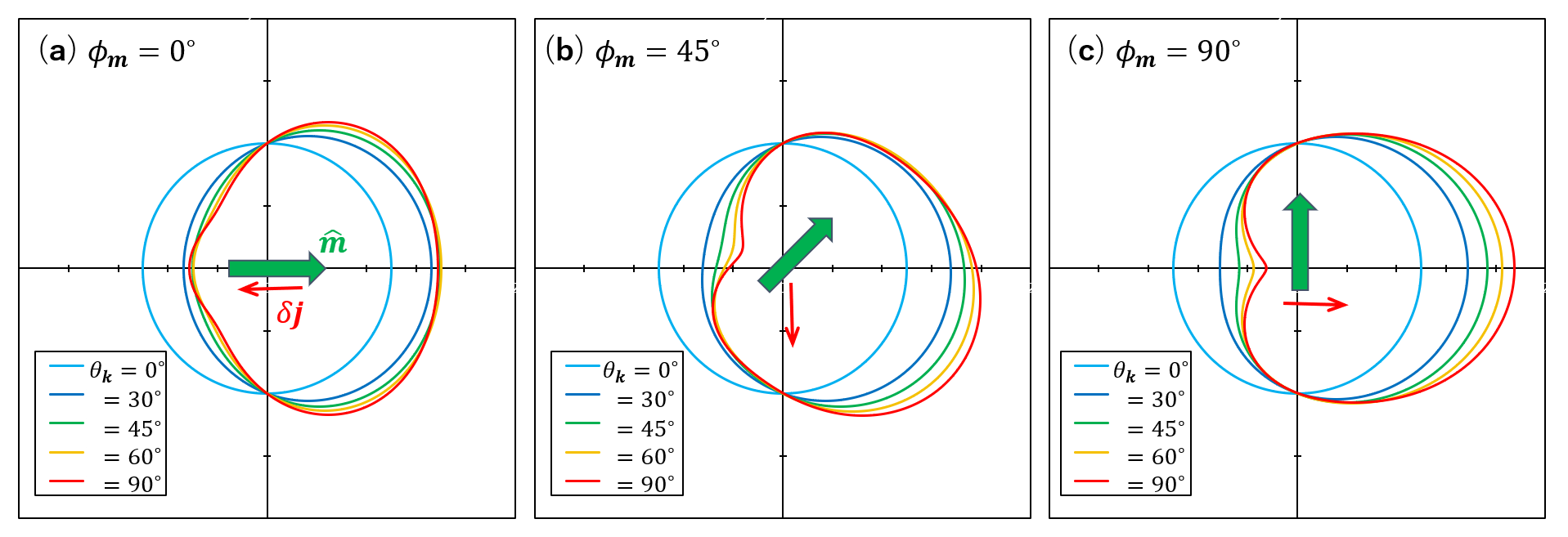}
	\caption{(Color online) Polar plot for the non-equilibirium charge distribution function $f_0$ with respect to $\phik$ for the different magnetization directions: (a) $\phim=0^\circ$, (b) $\phim=45^\circ$ and (c) $\phim=90^\circ$, where $\phik$ and $\phim$ are the azimuthal angles of $\unit{k}$ and $\unit{m}$ on the stationary coordinate system. Each line corresponds to the projection of $\thetak$ and the polar angle of $\unit{k}$. 
	The bold arrow represents $\unit{m}$, and the thin red arrow represents the anisotropic part of the charge current induced by the direction-dependent distribution.}
	\label{fig:Schema:AMRandPHE_phi}
\end{figure}

\newcommand{\feq}[1]{f_{\mathrm{eq},#1}}
To provide an intuitive understanding of both  ASF-SHE and ASF-MSHE, we constructed a kinetic view of these phenomena using the equations of motion based on the Boltzmann transport theory in the spin or space.
We define the $2 \times 2$ matrix of the non-equilibrium electron distribution function as
\begin{equation}\label{eq:Def:DistFunc}
\hat{f}(\boldsymbol{k})=\pauli{0}f_0(\boldsymbol{k}) +\unit{\sigma}\cdot\boldsymbol{f}(\boldsymbol{k}),
\end{equation}
where $f_0$ is the charge distribution, and $f_\mu$ is the spin distribution with spin $\mu=\theta,\phi,m$.
We assume a three-dimensional ferromagnetic electron-gas with a weak SOI, which leads to a spin-diagonal equilibrium distribution function with $\feq{\up}$ and $\feq{\dn}$, and start with the Boltzmann equation for a steady state with the relaxation time approximation\cite{Morawetz2015}
\begin{equation}\label{eq:Def:Boltzmann}
-\frac{eE_x v_x}{\hbar} \hat{f}'(\boldsymbol{k})=-\frac{1}{2}\{\hat{\tau}^{-1},\delta\hat{f}(\boldsymbol{k})\},
\end{equation}
with $\hat{f}'(\boldsymbol{k})=\partial \hat{f}(\boldsymbol{k})/\partial \varepsilon_{\up,\bm{k}}=\partial \hat{f}(\boldsymbol{k})/\partial \varepsilon_{\dn,\bm{k}}$.
 $\delta \hat{f}(\boldsymbol{k})=\hat{f}(\boldsymbol{k})-\hat{f}_{\mathrm{eq}}(\boldsymbol{k})$ denote the deviation from the equilibrium distribution. 
The relaxation time $\hat{\tau}$ is spin-dependent and is represented by a $2\times 2$ matrix.
Hereinafter, we assume the $\phi$-scan ($\thetam=\pi/2$) for the magnetization direction until the end of this section.
Using the results of Eqs. \eqref{eq:Approx:Tmatrix} and \eqref{eq:Approx:SingleSiteTmatrix}, we define a relaxation time which depends on the momentum direction $(\thetak, \phik)$ as well as the magnetization direction $\phim$, and obtain a linear response solution of Eq. \eqref{eq:Def:Boltzmann}
\begin{align}
\label{eq:Approx:FermiDist_0}
\delta f_{0}(\boldsymbol{k}) &\propto \left\{ h_0 +(\feq{\up}'(\boldsymbol{k})-\feq{\dn}'(\boldsymbol{k})) \sin^2 \thetak \cos^2(\phik-\phim)\right\}\sin\thetak \cos \phik,\\
\label{eq:Approx:FermiDist_3}
\delta f_{m}(\boldsymbol{k}) &\propto \left\{ h_z +(\feq{\up}'(\boldsymbol{k})+\feq{\dn}'(\boldsymbol{k}))  \sin^2 \thetak \cos^2(\phik-\phim)\right\}\sin\thetak \cos \phik , \\
\label{eq:Approx:FermiDist_1}
\delta f_{\theta}(\boldsymbol{k}) &\propto(\feq{\up}'(\boldsymbol{k})+\feq{\dn}'(\boldsymbol{k})) \sin^2\thetak \sin 2(\phik-\phim)\sin\thetak \cos \phik,\\
\label{eq:Approx:FermiDist_2}
\delta f_{\phi}(\boldsymbol{k}) &\propto(\feq{\up}'(\boldsymbol{k})-\feq{\dn}'(\boldsymbol{k}))  \sin^2\thetak \sin 2(\phik-\phim)\sin\thetak \cos \phik,
\end{align}
where the terms $h_0$ and $h_z$ are independent of $\unit{k}$ and $\unit{m}$, respectively.
The derivation is given in Appendix \ref{sec:Appendix:Boltzmann}.
First, we observe the momentum dependence of the charge distribution, $f_0(\boldsymbol{k})$, as plotted in Fig. \ref{fig:Schema:AMRandPHE_phi} (a)-(c).
Reflecting the s-d scattering nature of the relaxation time, $\delta f_0(\boldsymbol{k})$ shows an anisotropic distribution depending on the relative angle between the magnetization angle $\phim$ and momentum angle $\phik$.
Comparing $\phim=0$ and $\phim=\frac{\pi}{2}$ in Fig. \ref{fig:Schema:AMRandPHE_phi}, the distribution along the $x$-axis is different, and therefore the longitudinal current $j_x^0$ is modulated by the direction of magnetization.
This behavior can be understood as the AMR effect.
For $\phim=\frac{\pi}{4}$, the distribution is biased in the $y$-axis direction and provides finite transverse currents $j_y^0$, which is the PHE.

\begin{figure}
	\centering
	\includegraphics[width=\linewidth]{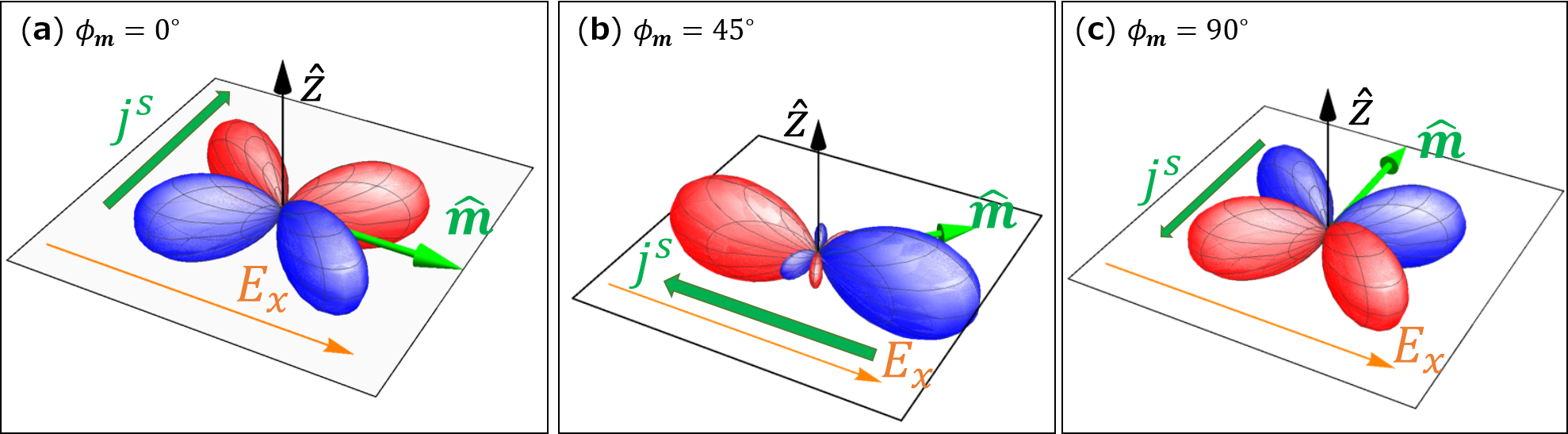}
	\caption{(Color online) Spherical plot for the isosurface of a non-equilibrium spin distribution function $\delta f_\theta$ with respect to $\phik$ for the different magnetization directions: (a) $\phim=0^\circ$, (b) $\phim=45^\circ$ and (c) $\phim=90^\circ$. 
		The colored area represents the sign of the distribution function, with red and blue denoting positive and negative signs, respectively.
		The $\up$ spins are more likely to flip to $\dn$ for the red area, and vice versa for the blue area, which corresponds to the illustration in Fig. \ref{fig:Schema:ASF}. 
		Note that $\delta f_\phi$ shows the same isosurface as $\delta f_\theta$.
	}
	\label{fig:Schema:ASHEandMSHE}
\end{figure}

Similarly, $\delta f_\theta(\boldsymbol{k})$ and $\delta f_\phi(\boldsymbol{k})$ also have momentum dependent distributions 
as shown in Fig. \ref{fig:Schema:ASHEandMSHE} (a)-(c), resulting in non-equilibrium spin-currents polarized along $\unit{\theta}$ or $\unit{\phi}$.
When $\phim=0$ or $\phim=\frac{\pi}{2}$, as shown in Fig. \ref{fig:Schema:ASHEandMSHE}, the finite transverse spin-currents $j_y^\theta, j_y^\phi$, which correspond to ASF-SHE/MSHE, exist.
Namely, the ASF-SHE/MSHE can be understood intuitively in a similar way to the PHE in terms of the anisotropic spin-dependent relaxation time. The ASF-SHE/MSHE is governed by spin-flip scattering, whereas the PHE is governed by spin-dependent momentum scattering.
When $\phim=\frac{\pi}{4}$, there are finite longitudinal spin currents $j_x^\theta, j_x^\phi$, which can be understood similarly to the AMR effect. 

Note that a non-equilibrium anisotropic spin distribution of this nature is not specific to the momentum- and magnetization-dependent relaxation time, as discussed here. When the Fermi surface has spin-momentum locking, a constant relaxation time can also generate a spin distribution and a resultant spin current, which can show the same magnetization dependence \cite{Zelezny2017}. 
Although both mechanisms can contribute to the spin current with the same symmetry, they show quantitatively different dependencies on parameters, reflecting their different microscopic origins. 
Therefore, we would obtain both contributions and their interference effect if we were to utilize both the spin-momentum locking Fermi surface and magnetization-dependent relaxation time.

\subsection{Numerical calculation}
\begin{figure}[h]
	\centering
	\includegraphics[width=0.7\linewidth]{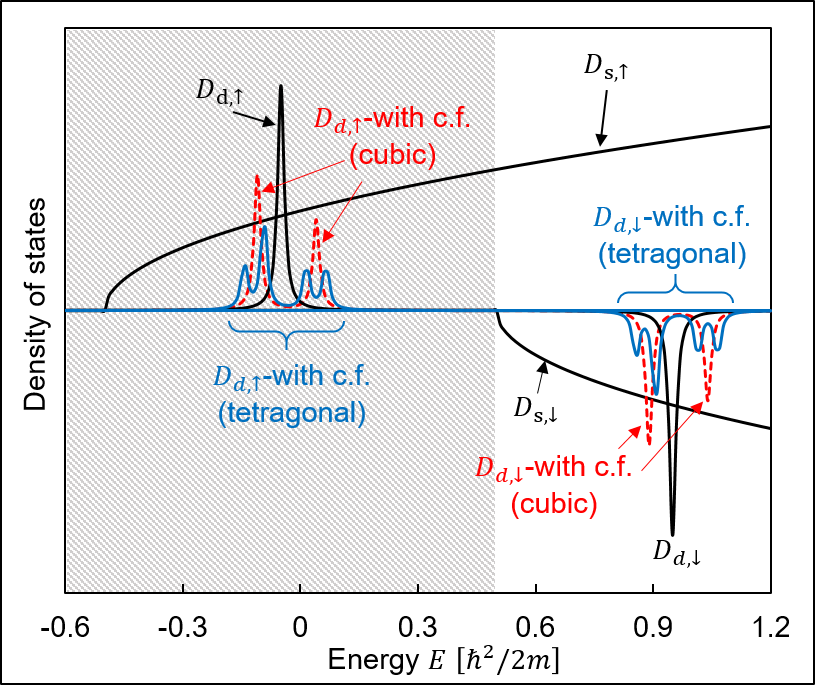}
	\caption{Schematics of the projected density of states (PDOS) of the conduction bands and the 3d level for $\Ed=0.45$ and $\Dels=\Deld=0.5$.
	The solid gray and solid black lines represent the PDOS of the conduction band and the PDOS of the 3d states for each spin, respectively. The dotted red and blue lines represent the PDOS of each splitting 3d level due to the cubic and tetragonal crystal fields, respectively. 
	The hatched area represents a half-metallic region, where a single spin conduction band is present.
	To improve visualization, the spin-orbit interaction is ignored, and the spectral widths are adjusted using different values from those in the numerical calculations.}
	\label{fig:Schema:DOS}
\end{figure}

\newcommand{\ASHA}{\tilde{\sigma}_{\perp}^{\mathrm{(even)}}}
\newcommand{\MSHA}{\tilde{\sigma}_{\perp}^{\mathrm{(odd)}}}
\newcommand{\SPHA}{\tilde{\sigma}_{\parallel}^{\mathrm{(odd)}}}
We numerically evaluated Eqs. \eqref{eq:Approx:KuboStreda1} and \eqref{eq:Approx:KuboStreda2} to test the validity of the perturbative analysis developed in the previous section and qualitatively investigate the dependence on several parameters.
In this section, we also include the crystal-field effects.
We set the parameters $\Dels=\Deld=0.5$ and $\xi=-0.025$ in units of $\hbar^2/2m$, which we consider to be reasonable for ferromagnetic metals.
We take the bottom of the conduction electron band as the reference point for the energy and set $\Ed=0.45$. The density of states (DOS) is shown in Fig. \ref{fig:Schema:DOS}.
For simplicity, we assume $V_\up(\kF{})=V_\dn(\kF{})=V$, and choose the parameters related to impurity scattering as $\nimp V^2=0.1$ and $\selfenergyss=\dwidth=0.1$.

\begin{figure}
	\centering
	\includegraphics[width=0.7\linewidth]{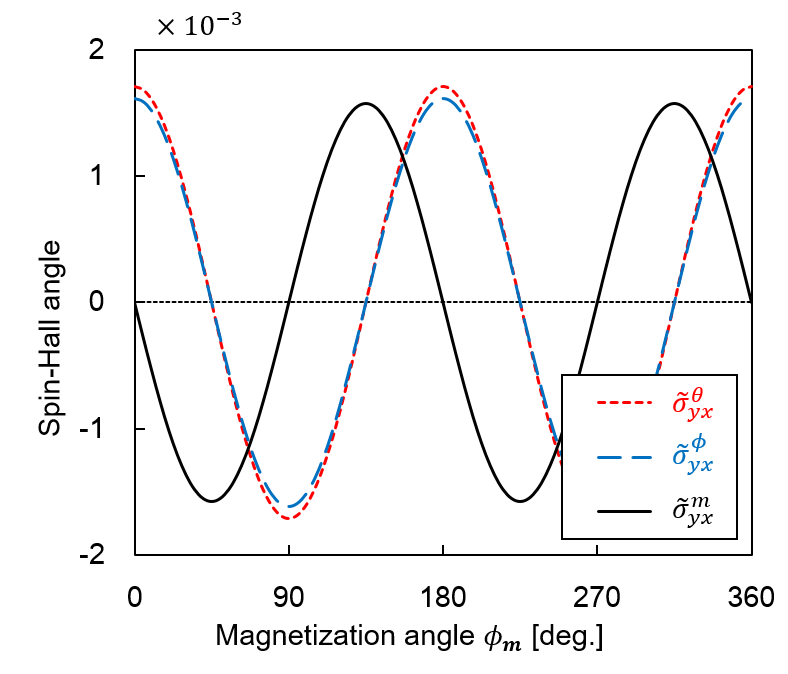}
	\caption{(Color online) Spin-Hall angles $\tilde{\sigma}^\mu_{yx}\equiv -(2e/\hbar)\sigma^\mu_{yx}/\sigma^0_{xx}$ as functions of the in-plane magnetization direction $\phi$.}
	\label{fig:Result:SHE_Phi_woCF}
\end{figure}

\begin{figure}[h]
	\centering
	\includegraphics[width=0.7\linewidth]{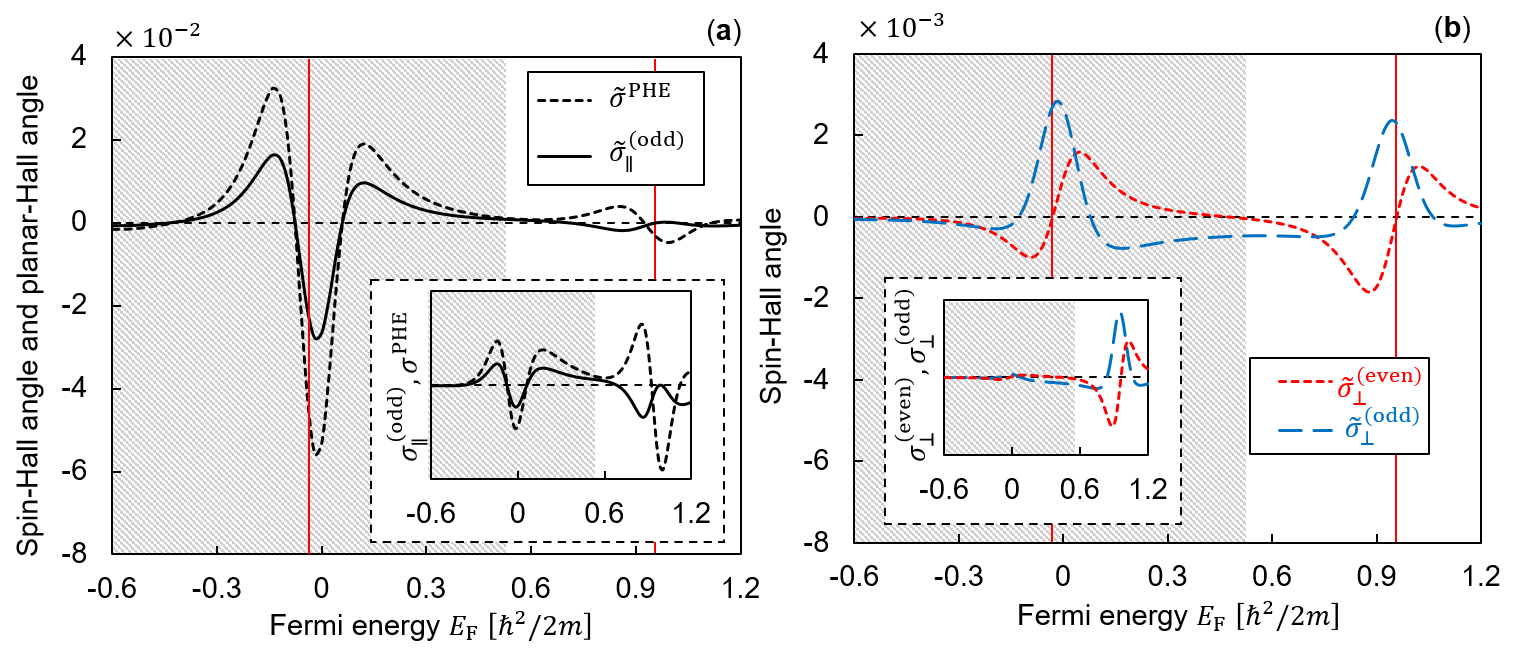}
	\caption{(Color online) Fermi energy dependences of the spin Hall angles for (a) $\sigmaSPHE$ and (b) $\sigmaASHE$, $\sigmaMSHE$. The vertical red line represents the position of each 3d level. The insets show the intensities of each spin conductivity before being normalized by the longitudinal conductivity.
	The planar-Hall angle is also plotted in (a).
	The hatched area represents the half-metallic region shown in Fig. \ref{fig:Schema:DOS}.}
	\label{fig:ResultFEwoCF}
\end{figure}

First, we show the dependence of the spin conductivities on the direction of magnetization and Fermi energy in the case of $\Delc=\DelT=0$, and then compare the results of the perturbative analysis.
Hereinafter, we consider the spin Hall angles (SHAs) $\tilde{\sigma}_{yx}^\mu \equiv -\frac{\hbar}{2e}\sigma_{yx}^\mu$ instead of $\sigma_{yx}^\mu$.
Fig. \ref{fig:Result:SHE_Phi_woCF} shows the dependences of the SHAs on the magnetization angle in the $\phi$-scan ($\thetam=\pi/2$) with $\EF=0.9$.
These results are consistent with the prediction of the perturbative analysis in Eqs. \eqref{eq:Result:SPHE}-\eqref{eq:Result:MSHE} because we find that the $\sigma_{yx}^\theta, \sigma_{yx}^\phi\propto\cos 2\phi$ and $\sigma_{yx}^m\propto \sin 2 \phi$.
The Fermi energy dependences of $\ASHA,\MSHA,\SPHA$, and the planar Hall angle (PHA) are shown in Fig. \ref{fig:ResultFEwoCF}(a) and (b). 
The results show that all of the plots have peaks because of resonant scattering in the vicinity of each impurity level. 
Referring to Eqs. \eqref{eq:Approx:SPHE}-\eqref{eq:Approx:MSHE}, it is clear that these resonance peaks reflect the spectral structure of Green's functions of each of the d-orbital states. 
In panel (a), the planar Hall angle (PHA) takes values of approximately $\lesssim2 \%$, which seems of a reasonable order of magnitude in a typical ferromagnetic metal, although we use a simplified model with roughly estimated parameters.
In the half-metallic region ($\EF\lesssim0.55$), $\sigmaSPHE$ is roughly proportional to the PHA, supporting the SP-PHE.
In the normal metallic region ($\EF \gtrsim 0.55$), the relationship between $\SPHA$ and PHA becomes more complex because the contribution from both spin bands is tangled.
$\SPHA$ tends to be enhanced in the half-metal region rather than in the normal metallic region. This is because $\sigma_{xx}$ in the half-metal region is smaller than that in the normal metal region. 
In contrast, $\sigmaSPHE$ maintains the same order of magnitude from the half-metal region to the normal metal region.
On the other hand, in panel (b), because $\ASHA$ and $\MSHA$ are governed by the mixing of $\Upsilon_\up$ and $\Upsilon_\dn$, little difference in magnitude exists between the half-metallic and normal metallic regions.
Note that this is the case for the SHA; however, the absolute value of the spin conductivity itself increases with increasing longitudinal conductivity, as shown in the inset of  Fig. \ref{fig:ResultFEwoCF}(b).
Comparing the behavior of $\ASHA$ and $\MSHA$ with Eqs. \eqref{eq:Approx:ASHE} and \eqref{eq:Approx:MSHE}, it becomes possible to obtain the characteristics of the real and the imaginary parts of the Green's function.

\begin{figure}
\centering
\includegraphics[width=0.7\linewidth]{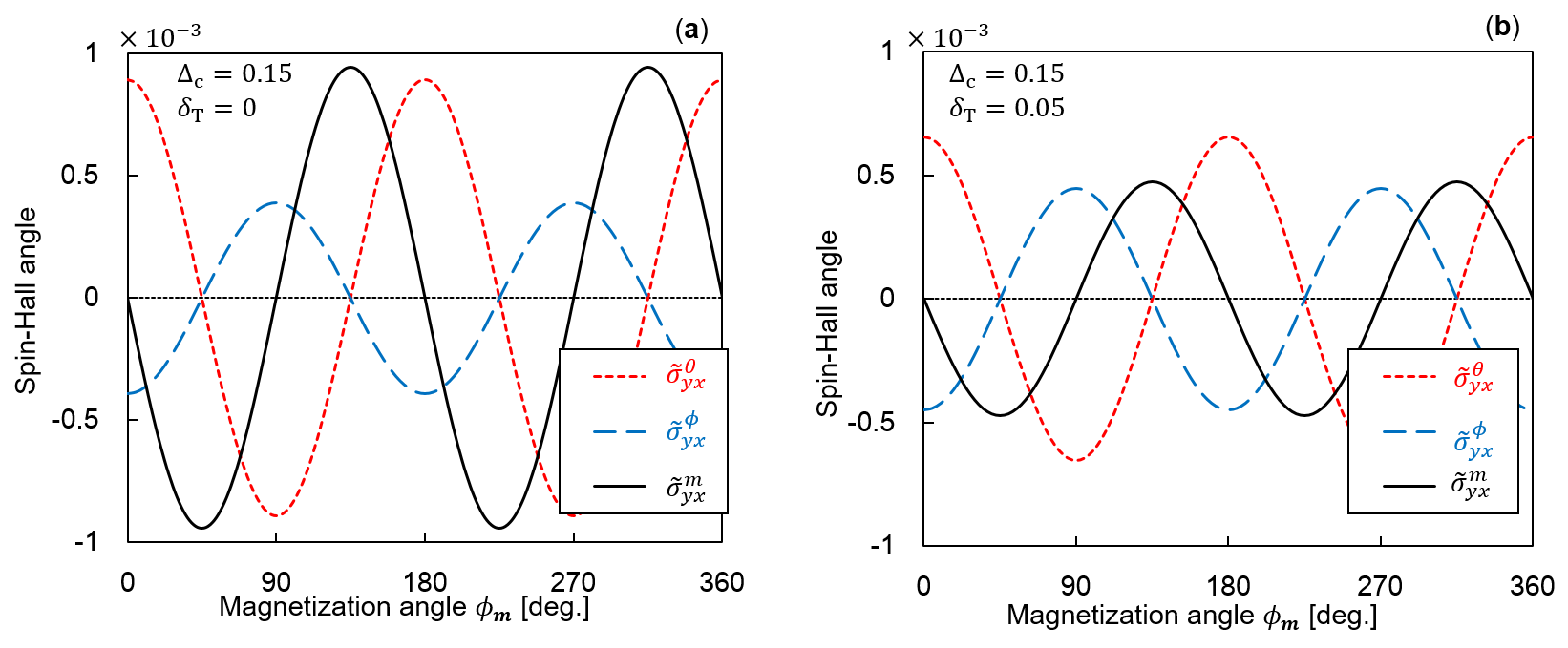}
\caption{(Color online) Spin-Hall angles $\tilde{\sigma}^\mu_{yx}\equiv -(2e/\hbar)\sigma^\mu_{yx}/\sigma^0_{xx}$ as functions of the in-plane magnetization direction $\phi$ with (a) the cubic field splitting $\Delc=0.15, \DelT=0.0$ and (b) the tetragonal field splitting $\Delc=0.15, \DelT=0.05$ .}
\label{fig:ResultPhiCF}
\end{figure}

Next, we present the calculation results by taking into account the cubic or tetragonal crystal field, either of which splits the 3d-orbital states, as shown in Fig. \ref{fig:Schema:DOS}. 
Fig. \ref{fig:ResultPhiCF}(a) and (b) show the magnetization angular dependences of SHAs under the cubic field ($\Delc=0.15,\DelT=0$) and the tetragonal field ($\Delc\neq0,\DelT\neq0$).
Compared to the results obtained without the crystal field (Fig. \ref{fig:Result:SHE_Phi_woCF}), the crystal fields change not only the angular dependency but the amplitudes in each component of the SHA.

\begin{figure}
\centering
\includegraphics[width=0.7\linewidth]{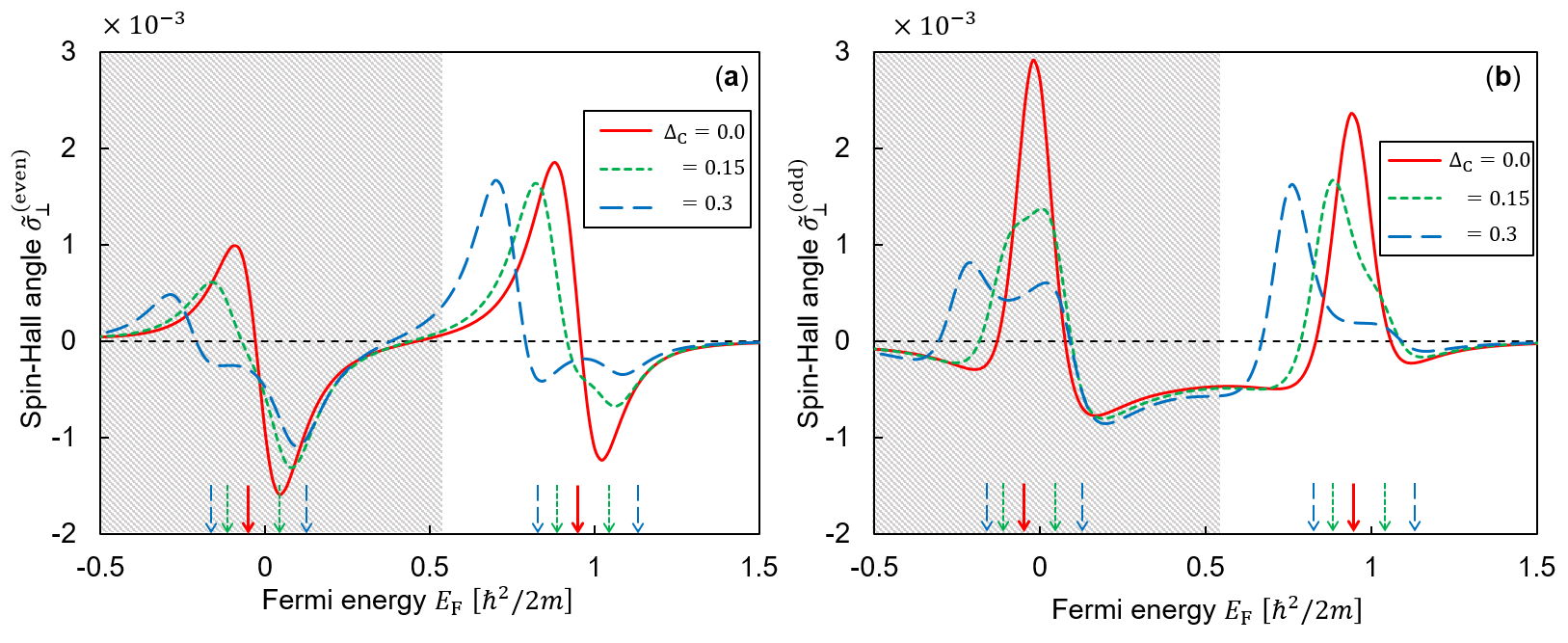}
\caption{(Color online) Fermi energy dependences of spin Hall angles for (a) $\sigmaASHE$ and (b)$\sigmaMSHE$ for various values of the cubic crystal field $\Delc=0.0,0.15,0.3$. The arrows at the bottom represent the positions of the corresponding energy levels.}
\label{fig:Result:EFwCubic}
\end{figure}

\begin{figure}
\centering
\includegraphics[width=0.7\linewidth]{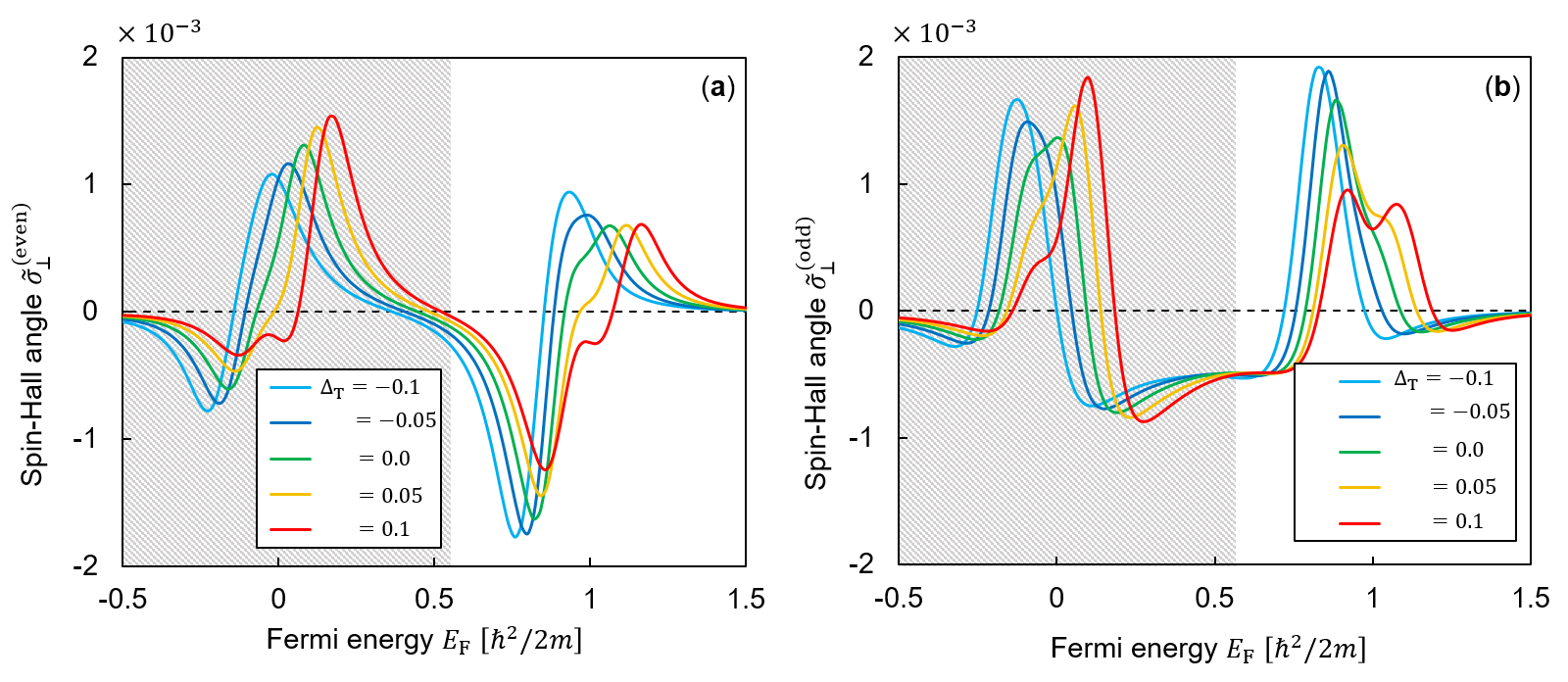}
\caption{(Color online) Fermi energy dependences of spin Hall angles for (a) $\sigmaASHE$ and (b)$\sigmaMSHE$ for various values of the tetragonal distortion $\DelT=-0.1,-0.05,0.0,0.05,0.1$ and $\Delc=0.15$ (fixed). }
\label{fig:Result:EFwTetra}
\end{figure}

Fig. \ref{fig:Result:EFwCubic} shows the Fermi energy dependences of $\ASHA$ and $\MSHA$ for various values of the cubic field.
In both results, the resonance peaks shift together with the splitting 3d-levels when $\Delc\neq0$, whereas the behavior around each peak is similar to that in the case of $\Delc=0$.
This trend also holds for the tetragonal crystal field.
Fig. \ref{fig:Result:EFwTetra} shows the Fermi energy dependences of $\ASHA$ and $\MSHA$ for various values of the tetragonal field.
Similar to Fig. \ref{fig:Result:EFwCubic}, the peaks shift together with the 3d-levels without any drastic change of the functions.
These results suggest that the cubic or tetragonal crystal field can modify the magnitude of the SHAs via the splitting of the 3d-states, but do not change the physical view of $\ASHA$ and $\MSHA$ that arise from the s-d scattering with anisotropic spin flipping.
Therefore, we expect both the ASF-SHE and ASF-MSHE to be robust against the crystal field effect and can appear in actual materials.

\section{Summary}
\label{sec:conclusion}
We presented a systematic investigation of the spin-dependent scattering mechanisms of the spin Hall effect (SHE) in a ferromagnetic metal and proposed a new mechanism for SHE.
We assumed the existence of s-d scattering in a ferromagnetic 3d alloy and represented it using a multi-orbital impurity Anderson model considering the spin-orbit interaction (SOI) and the crystal-field splitting of the d-orbital states.
The spin conductivities were formulated using microscopic transport theory based on the Kubo formula within the averaged T-matrix approximation for randomness.
To determine the physical aspect of the spin conductivity, we first performed an analytical derivation using a perturbation expansion with respect to the SOI up to the second order.
As a result, we obtained both contributions to the time-reversal even (SHE) and the time-reversal odd (magnetic SHE) from the anisotropic spin-flip scattering (ASF) process:
a scattering process that combines the anisotropic (spatially dependent) s-d hybridization and spin-flip by the SOI.
From a microscopic viewpoint, the ASF can be understood as the coupling between the momentum of the s-electron and its spin via the intermediate d-orbital of an impurity, which is a result of the combination of the SOI in the d-orbital and the orbital-selective s-d transition.
The spin current of both effects follows $\cos 2\phi$, where $\phi$ is the relative angle between the magnetization and the applied electric field, and their spin Hall resistivities approximately scale linearly with the longitudinal resistivity in the diffusive metal region.
For the analytical calculation, we disregarded the crystal-field splitting for simplicity.
In the kinetic view, the ASF scattering processes are responsible for a $\phi$-dependent spin-flip relaxation time, which induces an anisotropic non-equilibrium spin distribution and consequently generates a finite spin current.
This is similar to the planar Hall effect (PHE) in bulk ferromagnets, which arises from an anisotropic non-equilibrium charge distribution.
A distinctive feature of the ASF-SHE is that the spin current is polarized perpendicular to $\unit{m}$ and controllable by the magnetization direction.
We also performed numerical calculations to determine the influence of the crystal-field splitting.
The $\phi$ dependences and Fermi energy dependences were computed for different strengths of the crystal field.
As a result of their dependences on $\phi$, both the spin Hall angle (SHA) of the ASF-SHE and the ASF-MSHE were proportional to $\cos 2\phi$, in agreement with the perturbative results even with cubic or tetragonal crystal fields.
From the Fermi energy dependences, we found that the crystal fields appear to produce only energy shifts owing to the level splitting among d-states and do not suppress the spin current.
The results suggest that both the ASF-SHE and ASF-MSHE are robust against orbital splitting owing to the crystal fields; therefore, they can appear in actual materials.
Although both the ASF-SHE and ASF-MSHE have not yet been experimentally determined, we expect that a contribution from these effects can be involved in the SHE signals of precursive measurements in ferromagnetic metals.
If they occur, we expect their contribution to show linear scaling with the longitudinal charge conductivity and a two-fold ($\cos 2\phi$) dependence with respect to the relative angle between the applied electric field and the magnetization direction.

\section*{Acknowledgments}
This study was supported by the Center for Spintronics Research Network (CSRN).
Yuta Yahagi acknowledges support from GP-Spin at Tohoku University.

\appendix
\section{Longitudinal spin conduction}
\label{sec:Appendix:Longitudinal}
Although the generation of the transverse spin current resulting from the anisotropic spin-flip mechanism is mainly discussed in this work, this mechanism can also generate an additional contribution to the longitudinal spin current by the same processes as in Fig. \ref{fig:Diagram:ElementaryProcess}.
We obtain spin conductivities $\sigma_{xx}^\mu$
\begin{align}
\label{eq:Result:AMSC}
\sigma_{xx}^m=&\bar{\sigma}^{3}_{xx}
+\sigmaMSHE \{1+(1-\cos 2\thetam)\cos^2 \phim\}
,\\
\label{eq:Result:LASHE}
\sigma_{xx}^\theta=&-\sigmaMSHE \sin 2\thetam \cos^2 \phim
-\sigmaASHE\sin \thetam \sin 2\phim
,\\
\label{eq:Result:LMSHC}
\sigma_{xx}^\phi=&
\sigmaMSHE \sin \thetam \sin 2\phim
-\sigmaASHE \sin 2\thetam \cos^2 \phim
,
\end{align}
where $\bar{\sigma}^{3}_{xx}=\frac{\hbar}{2(-e)}(\sigma_\up-\sigma_\dn)$ is the spin current associated with the (non-relativistic) spin-polarized current.  
The first term corresponds to the modification of the spin-polarized current due to the AMR, analogous to Eq. \eqref{eq:Result:SPHE}, which describes the PHE-SC. 
The second and third terms can be understood as the longitudinal counterparts of Eqs. \eqref{eq:Result:ASHE} and \eqref{eq:Result:MSHE}.

\section{Derivation of spin-conductivity}
\label{sec:Appendix:Derivation_Perturbation}
We derive an analytical expression for spin-conductivity by using a perturbation theory. 
We start from the matrix T in Eqs. \eqref{eq:Approx:SingleSiteTmatrix}.
After taking the perturbation up to the 2nd order with respect to the SOI strength, the single-site T-matrix $\hat{t}_{\boldsymbol{k,k}}$ can be expanded as
\begin{align}
\hat{t}_{\boldsymbol{k,k}}&\simeq \hat{t}_0 +\hat{t}'_0+\hat{t}'_\parallel + \hat{t}'_\perp, \\
\hat{t}_0 &\equiv \frac{5}{4\pi}\begin{pmatrix}V_\up^2 \Upsilon_\up & 0 \\ 0& V_\dn^2 \Upsilon_\dn \end{pmatrix}, \\
\hat{t}'_0&\equiv \frac{15}{4\pi}\lambda^2(\Upsilon_\up+\Upsilon_\dn)\begin{pmatrix}V_\up^2 \Upsilon_\up^2&0\\0&V_\dn^2 \Upsilon_\dn^2\end{pmatrix},\\
\label{eq:Res:SE_parallel_krepres}
\hat{t}'_\parallel&\equiv-\frac{15}{4\pi}\lambda^2(\Upsilon_\up-\Upsilon_\dn)
\begin{pmatrix}V_\up^2 \Upsilon_\up^2&0\\0&-V_\dn^2 \Upsilon_\dn^2\end{pmatrix}\hat{k}_m^2,\\
\label{eq:Res:SE_perp_krepres}
\hat{t}'_\perp&\equiv -\frac{15}{4\pi}\lambda^2 V_\up V_\dn \Upsilon_\up \Upsilon_\dn(\Upsilon_\up-\Upsilon_\dn)
(\pauli{1}\hat{k}_\theta + \pauli{2}\hat{k}_\phi)\hat{k}_m,
\end{align}
where $\hat{t}_0$ is the nonperturbative term, and the remainder are the perturbed terms.
Among the perturbed terms, $\hat{t}'_0$ represents the isotropic parts with respect to the momentum vector $\boldsymbol{k}$, 
whereas $\hat{t}'_\parallel$ and $\hat{t}'_\perp$ represent the spin-diagonal and off-diagonal parts of the anisotropic ($\unit{k}$-dependent) terms, respectively.
By substituting these terms into Eq. \eqref{eq:Def:GreensFunction}, $\Gk$ are expanded as 
\begin{equation}\label{eq:Approx:PertubGF}
\hat{G}\simeq \hat{G}_0+\nimp \hat{G}_0(\hat{t}'_0+\hat{t}'_\parallel + \hat{t}'_\perp)\hat{G}_0,
\end{equation}
where $\hat{G}_0\equiv(E\pauli{0}-\hat{H}^{\mathrm{s}}-\nimp\hat{t}_0  \mp\i \selfenergyss)^{-1}$ is the nonperturbative term of Green's function.
We substitute this equation in Eq. \eqref{eq:Approx:KuboStreda1} and \eqref{eq:Approx:KuboStreda2}.
Here, using the isotropic shape of $\hat{G}_0$ in the momentum space, we carry out only the angular integration explicitly, and the remaining parts, including radial integrations, are substituted into the coefficients.
Then we can obtain Eqs. \eqref{eq:Result:SPHE}-\eqref{eq:Result:MSHE},
with the coefficients
\begin{align}
\sigmaSPHE=&\frac{15}{4\pi^3}\nimp e \left(\frac{\lambda\hbar^2}{m}\right)^2 \int_{0}^{\infty}dk k^4 \Re 
(\Upsilon_\up^+-\Upsilon_\dn^+) \\
&\quad \left\{
V_\up^2(\Upsilon_\up^+)^2 G_{0,\up}^+(G_{0,\up}^-)^2  -V_\dn^2(\Upsilon_\dn^+)^2 G_{0,\dn}^+(G_{0,\dn}^-)^2 
\right\}+( \cdots), \label{eq:Appendix:Coeff_odd_para} \\
\sigmaASHE=&-\frac{15}{4\pi^3}\nimp e \left(\frac{\lambda\hbar^2}{m}\right)^2 \int_{0}^{\infty}dk k^4 \Im 
V_\up V_\dn \Upsilon_\up^+\Upsilon_\dn^+ G_{0,\up}^+G_{0,\dn}^+
\left(
G_{0,\up}^- - G_{0,\dn}^-
\right) +(\cdots), \\
\sigmaMSHE=&\frac{15}{4\pi^3}\nimp e \left(\frac{\lambda\hbar^2}{m}\right)^2 \int_{0}^{\infty}dk k^4 \Re
V_\up V_\dn \Upsilon_\up^+\Upsilon_\dn^+ G_{0,\up}^+G_{0,\dn}^+
\left(
G_{0,\up}^- + G_{0,\dn}^-
\right) +(\cdots), \label{eq:Appendix:Coeff_odd_perp} 
\end{align}
where $G_{0,\sigma}^\pm=[\hat{G}_0]_{\sigma,\sigma}$ and $(\cdots)$ denote the terms including $(G_{0,\sigma}^\pm)^3$, whose contribution can be neglected in a diffusive metal region.

Because we consider dissipative conduction in metallic bands, we can neglect $(\cdots)$ in Eqs. \eqref{eq:Appendix:Coeff_odd_para}-\eqref{eq:Appendix:Coeff_odd_perp}.
Assuming $\selfenergyss=\frac{\hbar}{2\tau_0} \ll \EF$ and using the relations $G_{0,\sigma}^+(k)G_{0,\sigma}^-(k)\simeq \frac{2\pi \tau_\sigma}{\hbar}\delta(\EF-E_{k\sigma})$, $G_{0,\sigma}^+(\kF{\sigma})\simeq -\i \frac{\hbar}{\tau_0}$, and $G_{0,\sigma}^+(\kF{\bar{\sigma}})\simeq \sigma \frac{1}{2\Dels}$ with Fermi momentum of spin-$\sigma$ band $\kF{\sigma}$, we can derive Eqs. \eqref{eq:Approx:SPHE}-\eqref{eq:Approx:MSHE} from Eqs. \eqref{eq:Appendix:Coeff_odd_para}-\eqref{eq:Appendix:Coeff_odd_perp}.

\section{Spin conductivity in the stationary Cartesian coordinate system}
\label{sec:Appendix:Stationary}
The spin conductivities in the stationary Cartesian coordinate system are obtained directly by the following transformation:
\begin{equation}
\begin{pmatrix}
\sigma_{ij}^x\\
\sigma_{ij}^y\\
\sigma_{ij}^z
\end{pmatrix}
=\begin{pmatrix}
\cos \thetam \cos \phim& -\sin \phim & \sin \thetam \cos \phim \\
\cos \thetam \sin \phim & \cos \phim & \sin \thetam \sin \phim\\
-\sin \thetam & 0 & \cos \thetam
\end{pmatrix}\begin{pmatrix}
\sigma_{ij}^\theta\\
\sigma_{ij}^\phi\\
\sigma_{ij}^m
\end{pmatrix}.
\end{equation}
For example, in the case of the $\phi$-scan ($\thetam=\frac{\pi}{2}$), each term of $\sigma_{yx}^i,\quad(i=x,y,z)$ is expressed as
\begin{equation}
\begin{split}
\sigma_{yx}^x=&-\sigmaMSHE \cos 2\phim\sin\phim -\sigmaSPHE \sin 2\phim \cos \phim,\\
\sigma_{yx}^y=&\sigmaMSHE \cos 2\phim\cos\phim -\sigmaSPHE \sin 2\phim \sin \phim,\\
\sigma_{yx}^z=&\sigmaASHE \cos 2\phim.
\end{split}
\end{equation}
In the case of the $\theta$-scan ($\phim=0$) and $\theta'$-scan ($\phi=\frac{\pi}{2}$), we obtain
\begin{equation}
\begin{split}
\sigma_{yx}^x=&-\sigmaASHE \sin \thetam \cos \thetam,\\
\sigma_{yx}^y=&\sigmaMSHE\sin \thetam,\\
\sigma_{yx}^z=&\sigmaASHE \sin \thetam \sin \thetam,
\end{split}
\end{equation}
and
\begin{equation}
\begin{split}
\sigma_{yx}^x=&\sigmaMSHE \sin \thetam,\\
\sigma_{yx}^y=&\sigmaASHE\cos \thetam \sin \thetam,\\
\sigma_{yx}^z=&-\sigmaASHE \sin \thetam \sin \thetam.
\end{split}
\end{equation}

\section{Non-equilibrium spin distribution function from Boltzmann equation}
\label{sec:Appendix:Boltzmann}
In this section, we derive Eqs. \eqref{eq:Approx:FermiDist_0}-\eqref{eq:Approx:FermiDist_3} from Eq. \eqref{eq:Def:Boltzmann}.
First, we assume the conservation of the spin angular momentum in the stationary state, which means that
\begin{equation}\label{eq:Appendix:SpinConservation}
p_{\up\dn}f_{\dn\dn}+p_{\up\up}f_{\up\dn}=f_{\up\up}p_{\up\dn}+f_{\up\dn}p_{\dn\dn},
\end{equation}
with 
\begin{equation}
\begin{pmatrix}
f_{\up\up} & f_{\up\dn} \\ f_{\dn\up}  & f_{\dn\dn}
\end{pmatrix}=\hat{f}, \quad 
\begin{pmatrix}
p_{\up\up} & p_{\up\dn} \\ p_{\dn\up} & p_{\dn\dn}
\end{pmatrix}=\hat{\tau}^{-1}.
\end{equation}
The left-hand side of Eq. \eqref{eq:Appendix:SpinConservation} represents the probability of gaining $+\frac{\hbar}{2}$ angular momentum on average,
whereas the right-hand-side represents the probability of losing.
When the SOI is sufficiently small, we can ignore the second-order terms of $f_\perp$ and $p_\perp$, and Eq. \eqref{eq:Appendix:SpinConservation} leads $[\hat{p}, \hat{f}]=\hat{p}\hat{f}-\hat{f}\hat{p}\simeq0$.
Using this, we can rewrite Eq. \eqref{eq:Def:Boltzmann} to
\begin{equation}
\delta \hat{f}=\frac{eE_x v_x}{\hbar}\hat{\tau}\hat{f}'.
\end{equation}
The relaxation time can be derived from $\braket{\hat{T}_{\boldsymbol{k,k'}}(E)}_{\mathrm{conf}}$ given in Eq. \eqref{eq:Approx:Tmatrix} and \eqref{eq:Approx:SingleSiteTmatrix} with Fermi's golden rule
\begin{equation}
\label{eq:Def:SpinDependentLifetime}
\hat{\tau}(\unit{k},\unit{m})=\begin{pmatrix}
\bar{\tau}_{\up\up}+\tau'_{\up\up} \sin^2 \thetak \cos^2(\phik-\phim)  &
\tau'_{\up\dn}\sin^2\thetak \sin 2(\phik-\phim)\\
\tau'_{\dn\up} \sin^2\thetak \sin 2(\phik-\phim)&
\bar{\tau}_{\dn\dn} - \tau'_{\dn\dn} \sin^2 \thetak \cos^2(\phik-\phim)
\end{pmatrix},
\end{equation}
where $\tau'_{\sigma\sigma'}$ is the coefficient of the magnetization-dependent terms, and $\bar{\tau}_{\sigma\sigma}$ is the coefficient of the magnetization-independent terms.
By substituting this into Eq.\eqref{eq:Def:Boltzmann} and using $v_x\propto k_x \propto \sin\thetak \cos\phik$, we can obtain Eqs  \eqref{eq:Approx:FermiDist_0}-\eqref{eq:Approx:FermiDist_3}.

\bibliography{MSHE_2020_ref.bib}
%\printbibliography

\newpage
\section*{Figures}

\end{document}